\documentclass[10pt]{iopart}
\usepackage{citesort}
\newcommand{\ket}[1]{\left|{#1}\right\rangle}
\newcommand{\bra}[1]{\left\langle{#1}\right|}
\newcommand{\aver}[1]{\left\langle{#1}\right\rangle}

\newcommand{\modu}[1]{\left|{#1}\right|}

\newcommand{\bea}{\begin{eqnarray}}
\newcommand{\eea}{\end{eqnarray}}
\usepackage{graphicx}
\usepackage{epstopdf}
\usepackage{color}
\usepackage{epsfig}
\usepackage{cite}
\usepackage{amssymb}
\begin{document}

\title[Homodyne nonclassical area as a nonclassicality indicator ]{Homodyne nonclassical area as a nonclassicality indicator}
\author{M. Rohith$^{1,2,}\footnote{\ead{rohith.manayil@gmail.com}}$, S. Kannan$^3$, and C. Sudheesh$^4$}
\address{$^1$ Department of Physics, Government College,  Malappuram 676 509, University of Calicut, India.} 
\address{$^2$  P. G. \& Research Department of Physics, Government Arts and Science College, Kozhikode 673 018, University of Calicut, India.}
\address{$^3$ ISRO Inertial Systems Unit, Thiruvananthapuram, 695 013, India.}
\address{$^4$ Department of Physics, Indian Institute of Space Science and Technology, Thiruvananthapuram, 695 547, India.}

\vspace{10pt}
\begin{indented}
\item[]\today
\end{indented}

\begin{abstract}
We propose a legitimate and easily computable nonclassicality indicator for the states of electromagnetic field based on the standard deviation in the measurement of the homodyne rotated quadrature operator. The proposed nonclassicality indicator is the nonclassical area projected by the optical tomogram of the quantum state of light on the optical tomographic plane. If the nonclassical area projected by the optical tomogram of a quantum state is greater than zero, the state is  nonclassical, and the area is zero for the pure classical state. It is also noted that the nonclassical area of a quantum state increases with an increase in the strength of nonclassicality inducing operations on the state such as the squeezing, photon addition, etc. We have tested the validity of the nonclassical area measure by calculating the same for certain well-known nonclassical states and found that essential features of the nonclassicality shown by the states are captured in the nonclassical area. We have also shown that the nonclassical area is robust against environment-induced decoherence of the states.  
Nonclassical area projected by the optical tomogram of a quantum state of light is experimentally tractable using the balanced homodyne detection of the quadrature operator of the field, avoiding the reconstruction of the density matrix or the quasiprobability distribution of the state.
\end{abstract}

%
\vspace{2pc}

\submitto{\JPB}
%
%
\ioptwocol

\section{Introduction}
Properties of the coherent state of the radiation field can have analogous descriptions in the classical electrodynamics and are considered to be the most classical state of the field \cite{Glauber1963}. In contrast, the quantum states of the field which possesses certain exotic features that cannot have classical correspondence are referred to as the nonclassical states of light \cite{Dodonov2002}. Precisely, an arbitrary quantum state of light is said to be nonclassical if the Glauber-Sudarshan $P$-function of the state is highly singular or takes negative values somewhere in the phase space \cite{Glauber1963,Sudarshan1963}. Nonclassical states have received great theoretical attention as well as experiment interest over the past century primarily because of their use in technological applications ranging from gravitational wave detectors to quantum information protocols. Numerous varieties of nonclassical states have been experimentally prepared and were characterized by the optical homodyne tomography \cite{Lvovsky2009}. Significant theoretical efforts have been taken to study the various features shown by the nonclassical states of light \cite{Dodonov2002}. One can say that the degree of nonclassicality of a state is directly related to the nonclassical features. Therefore, the amount of nonclassicality or quantumness of the state is a primary parameter to be known before using a nonclassical state for technological applications.

There are several theoretical probes to identify a nonclassical state, and various measures have been proposed to quantify the degree of nonclassicality associated with an arbitrary quantum state of light. Mandel's $q$-parameter was used to characterize the deviation of photon-number statistics of the nonclassical states from the Poissonian photon-number statistics shown by the coherent state of the electromagnetic field \cite{Mandel1979}. The first attempt to quantify the degree of nonclassicality of a quantum state of light was proposed in terms of the nonclassical distance, which is the trace norm distance between the state and the set of all classical states \cite{Hillery1987}. Hereafter a variety of distance-based measures have been proposed based on the different types of metrics defining the distance \cite{Dodonov2000,Wunshe2001,Dodonov2003,Marian2002,Marian2004,Nair2017}. Since the evaluation of these measures requires optimization over the set of all classical states, distance-based measures are not easily computable.  

The entanglement potential of the state is another quantifier of the nonclassicality of state, which measures the entanglement created by a beam splitter in terms of the logarithmic negativity \cite{Asboth2005,Vidal2002}. A quantifier of nonclassicality and entanglement related to the number of quantum superpositions of classical states was also introduced \cite{Vogel2014}. Quasiprobability distributions and the characteristic function of the field contain information about the degree of nonclassicality of the state.  A nonclassicality indicator based on the volume of the negative part of the Wigner function in phase space was reported \cite{Kenfack2004}.
Since the squeezed states of light have a positive definite Wigner function \cite{Hudson1974,Kenfack2004}, the nonclassicality of such states cannot be characterized by the negativity of the Wigner function. Convolution transformation between Glauber-Sudarshan $P$-function and Husimi $Q$ function was used to define a universal nonclassicality measure called nonclassical depth \cite{Lee1991}. It can be viewed as the amount of thermal noise required to destroy whatever the nonclassical effects present in the quantum state. Recently, the characteristic function of the state has been used to quantify the nonclassicality of the state of light \cite{Ryl2017}. An operational measure of nonclassicality based on the negativity of an observable whose classical counterpart is positive semidefinite was introduced \cite{Gehrke2012}. Apart from the previously mentioned quantifiers, Schmidt-number witnesses have also been used to express the nonclassicality of the states \cite{Terhal2000,Sanpera2001,Mraz2014}.

Nonclassicality measures mentioned so far are not directly related to the experiment because the experimental estimation of these measures requires the knowledge of the density matrix or the quasiprobability distribution of the state, which is not a directly measurable quantity. One way to experimentally determine the density matrix or the quasiprobability distribution is using the optical homodyne tomography \cite{Lvovsky2009}. In this technique, a series of homodyne measurements of the rotated quadrature operator on an ensemble of identically prepared states generates an optical tomogram of the state \cite{Vogel1989,Smithey1993,Leonhardt1997}. The optical tomogram of the state is a primary object characterizing the state of light. It contains all the information about the state, including the amount of nonclassicality contained in the field. This optical tomogram is used further to reconstruct the density matrix or the quasiprobability distributions of the quantum state by numerical methods. The systematic and statistical errors associated with measurement may propagate during the reconstruction process and lead to the loss of information contained in the state. It was shown that the properties of a quantum state of light could be inferred directly from the optical tomogram of the state \cite{Bellini2012}. 

In this work, we have used the optical tomogram of the state to find a nonclassicality indicator for the quantum states of an electromagnetic field. This paper aims to introduce an easily computable nonclassicality indicator that is directly measurable using homodyne optical tomography. The proposed nonclassicality indicator is based on the standard deviation in measuring the homodyne rotated quadrature operator in a quantum state. It can be considered as an effective area spanned by the optical tomogram of the state on the optical tomographic plane. The only classical pure states are the coherent states for which the effective area projected by the optical tomograms is a constant. For the nonclassical states, the effective area is always greater than the value corresponding to classical states, and the difference between these two values is defined as the nonclassical area.
To check the validity of the nonclassical area measure, we have calculated it for certain well-known quantum states and analyzed whether the nonclassical area captures the essential features of the nonclassicality shown by the states. We have analytically evaluated the nonclassical area for the states obtained by the superposition of coherent states and the states obtained by the nonclassicality inducing operations such as photon addition and squeezing action on the coherent state and vacuum state.  

The robustness of the nonclassical area measure is tested by studying the effect of environment-induced decoherence of the states on the nonclassical area. We have used the zero temperature master equation to model the amplitude decay of the states due to interaction with the external environment and studied the time evolution of the nonclassical area for various initial states. The rest of the manuscript  is organized as follows. Section~\ref{sec2} gives a brief overview of the optical tomographic representation of the quantum state of an electromagnetic field. A systematic derivation of the nonclassical area indicator of nonclassicality of a single-mode field based on the standard deviation of the rotated quadrature operator is given in section~\ref{sec3}. Section~\ref{sec3} also describes the calculations of the nonclassical area spanned by the optical tomogram of various single-mode nonclassical states and the effect of decoherence on the nonclassical area. The definition of nonclassical area for a two-mode field and the explicit calculations of the same for various two-mode states are given in section~\ref{sec4}. The effect of environment-induced decoherence of the two-mode state on the nonclassical area is also discussed. Section~\ref{sec5} generalizes the results obtained for single-mode and two-mode state of the field to the case of a generic $p$-mode state of the electromagnetic field. Section~\ref{sec6} summarizes the main results of the paper.

\section{Optical tomographic representation}\label{sec2}
The homodyne rotated quadrature operator for a single-mode electromagnetic field is given by
\begin{equation}
\mathbb{X}_{\theta}= \frac{1}{\sqrt{2}}\left(a\, e^{-i\theta}+a^\dag  e^{i\theta}\right),\label{quadratureSingle}
\end{equation}
where $\theta$ represents the phase of the local oscillator in homodyne detection arrangement ($\theta\in \left[0,2\pi\right]$), and  $a$ and $a^\dag$ are the ladder operators of the single-mode field, respectively. The optical tomogram of an arbitrary state $\ket{\psi}$ is defined  as the probability distribution of the rotated quadrature operator $\mathbb{X}_{\theta}$ in the state $\ket{\psi}$. If $\ket{X_{\theta},\theta}$ is the eigenvector of the Hermitian operator $\mathbb{X}_{\theta}$ with eigenvalue $X_{\theta}$ \cite{Barnett1997}, the optical tomogram a quantum state with density matrix $\rho$ is given by \cite{Vogel1989}:
\begin{eqnarray}
\omega\left(X_{\theta},\theta\right)=\bra{X_{\theta},\theta}\rho\ket{X_{\theta},\theta}.
\label{opt_tomo_def}
\end{eqnarray}
In terms of the creation operator, the state $\ket{X_{\theta},\theta}$ can be written as 
\bea
\ket{X_{\theta},\theta}=\frac{1}{\pi^{1/4}} \exp\left[-\frac{{X_{\theta}}^2}{2}-\frac{1}{2} e^{i\,2\theta} {a^\dag}^2+\sqrt{2}\, e^{i\,\theta} X_{\theta}\, a^\dag\right]\ket{0},\nonumber\\
\eea
where $\ket{0}$ is the single-mode vacuum state. For a pure state with wave vector $\ket{\psi}$, the equation~(\ref{opt_tomo_def}) reduces to 
 \begin{eqnarray}
\omega(X_{\theta},\theta)=\modu{\bra{X_{\theta},\theta}\psi\rangle}^2,\label{opt_tomo_def_purestate}
\end{eqnarray}
where $\bra{X_{\theta},\theta}\psi\rangle$ is the quadrature representation of the state $\ket{\psi}$.  
The normalization condition of the optical tomogram $\omega (X_{\theta},\theta)$ is given by
\bea
\int_{-\infty}^{\infty} dX_\theta\, \omega (X_{\theta},\theta)=1.
\eea
The optical tomogram is having the following symmetry property:
\begin{equation}
\omega (X_{\theta},\theta+\pi)=\omega (-X_{\theta},\theta).\label{symmetry}
\end{equation}
The $n^{\rm th}$ moments of the operator $\mathbb{X}_\theta$ in an arbitrary quantum state can be calculated using the optical tomogram $\omega\left(X_\theta,\theta\right)$ of the corresponding state as
\begin{equation}
\aver{{\mathbb{X}_\theta^n}}=\int_{-\infty}^{\infty} dX_\theta \, {X_\theta^n}\, \omega\left(X_\theta,\theta\right). \label{moments}
\end{equation}
A plane with $X_\theta$- and $\theta$-axes defined is used to visualize the nonclassical features associated with the quantum states of light \cite{Rohith2015,Rohith2016,Sharmila2017}.

The above formalism can be extended to the case of a multimode field. For a $p$-mode electromagnetic field, the homodyne rotated quadrature can be written as
\begin{equation}
\mathbb{X}_{\theta_1,\,\theta_2,\,  \dots,\, \theta_p}=\frac{1}{\sqrt{p}} \sum_{i=1}^{p} \mathbb{X}_{\theta_i},\label{quadratureMultimode}
\end{equation}
where $\mathbb{X}_{\theta_i}$'s are the rotated quadrature operator for the individual modes as given in equation~(\ref{quadratureSingle}). The operator $\mathbb{X}_{\theta_1,\,\theta_2,\,  \dots,\, \theta_p}$ is defined in such way that it obeys the commutation relation given by $
\left[\mathbb{X}_{\theta_1,\,\theta_2,\,  \dots,\, \theta_p}, \mathbb{X}_{\theta_{1+\pi/2},\,\theta_{2+\pi/2},\,  \dots,\, \theta_{p+\pi/2}}\right]=i$. The optical tomogram of an $p$-mode state with density matrix $\rho$ can be defined as
\begin{eqnarray}
&\omega &\left(X_{\theta_1},\,\theta_1;\, X_{\theta_2},\,\theta_2;\, \dots;\,X_{\theta_p},\theta_p\right)=\nonumber\\
&&\bra{X_{\theta_1},\,\theta_1;\, X_{\theta_2},\,\theta_2;\, \dots;\,X_{\theta_p},\theta_p}\rho\left|X_{\theta_1},\,\theta_1;\, X_{\theta_2},\,\theta_2;\,\right.\nonumber\\
&&\left. \dots;\,X_{\theta_p},\theta_p\right>,
\label{opt_tomo_def_multimode}
\end{eqnarray}
where $\ket{X_{\theta_1},\,\theta_1;\, X_{\theta_2},\,\theta_2;\, \dots\,X_{\theta_p},\theta_p}$ is the eigenket of the $p$-mode rotated quadrature operator $\mathbb{X}_{\theta_1,\,\theta_2,\,  \dots,\, \theta_p}$ with eigenvalue $X_{\theta_1,\,\theta_2,\, \cdots,\,\theta_p}$. The $n^{\rm th}$ moment of the operator $\mathbb{X}_{\theta_1,\,\theta_2,\, \dots,\, \theta_p}$ for a generic $p$-mode state now becomes
\begin{eqnarray}
\aver{{\mathbb{X}_{\theta_1,\,\theta_2,\, \dots,\, \theta_p}^n}}&=&\int_{-\infty}^{\infty} dX_{\theta_1} \,dX_{\theta_2} \,\cdots \,dX_{\theta_p} \, {X_{\theta_1,\,\theta_2,\, \dots\,\,\theta_p}^n}\nonumber\\
&&\omega\left(X_{\theta_1},\,\theta_1;\, X_{\theta_2},\,\theta_2;\, \dots;\,X_{\theta_p},\theta_p\right). \label{moments-p-mode}
\end{eqnarray}
Optical tomograms of variety of nonclassical states have been theoretically investigated \cite{Bazrafkan2003,Filippov2011,Korennoy2011,Miranowicz2014,Rohith2015,Rohith2016}. Recently, it has been shown that the signatures of nonclassical effects such as quantum wave packet revivals, quadrature squeezing, and  entanglement can be captured in the optical tomogram of the state \cite{Rohith2015,Rohith2016,Sharmila2017}. Quantitative estimation of the degree of nonclassicality of an arbitrary quantum state of the light in terms of the optical tomogram of the state has not been reported so far.  

\section{Nonclassical area for a single-mode field}\label{sec3}
Consider an arbitrary single-mode quantum state $\ket{\psi}$ whose optical tomogram is given by $\omega\left(X_{\theta},\theta\right)$.  We use the amount of quadrature fluctuation in the measurement of the rotated quadrature operator to introduce a nonclassicality indicator associated with the state $\ket{\psi}$. Specifically, the quantity of interest is the standard deviation in the measurement of the rotated quadrature operator $\mathbb{X}_\theta$ in the state $\ket{\psi}$, defined as
\begin{equation}
\Delta X_\theta=\sqrt{\aver{{\mathbb{X}_\theta^2}}-{\aver{\mathbb{X}_\theta}}^2}.
\end{equation}
The standard deviation $\Delta X_\theta$ can be evaluated for any arbitrary state using the optical tomogram of the corresponding state with the help of equation~(\ref{moments}).  The standard deviation in the measurement of $\mathbb{X}_\theta$ reflects the amount of spread of optical tomogram of the state $\ket{\psi}$ on the optical tomographic plane for a particular $\theta$.

Measurements of two rotated quadrature operators with $\theta^\prime$ values differ by $\pi/2$ are limited by the Heisenberg uncertainty relation 
\begin{equation}
\Delta X_{\theta^\prime} \Delta X_{\theta^\prime+\pi/2} \geq \frac{1}{2},\label{uncertainty}
\end{equation}
as they are conjugate variables. In the expression~(\ref{uncertainty}) and  the rest of this manuscript, we have taken $\hbar=1$. A schematic diagram of the distribution of standard deviation $\Delta X_{\theta}$ with respect to $\theta$  (shaded region in the optical tomographic plane) is given in figure~\ref{schematic}.  The uncertainty relation [given in equation~(\ref{uncertainty})] equalizes for all $\theta$ values only for the classical  states of the electromagnetic field. Hence, it turns out that the area projected the distribution of standard deviation $\Delta X_{\theta}$ on the optical tomographic plane (the shaded region) will have a lower bound for the pure classical states of the field. 

Let us evaluate the lower bound of the area projected by the distribution of standard deviation $\Delta X_{\theta}$  on the optical tomographic plane for a pure single-mode classical state. We divide the optical tomographic plane into horizontal strips having minimal width $\Delta \theta$ so that the value of $\Delta X_\theta$ is constant throughout the strip. The symmetry property of the optical tomogram given in equation~(\ref{symmetry}) allows us to confirm that the area spanned by the standard deviation distribution on the optical tomographic plane in the region between $\theta=0$ to $\theta=2\pi$ is two times the area of spanned by the same in the region between $\theta=0$ to $\theta=\pi$. Again, we divide the region between $\theta=0$ to $\theta=\pi$ into two  halves. One between $\theta=0$ to $\theta=\pi/2$ and other between $\theta=\pi/2$ to $\theta=\pi$. For each horizontal strip in the first half, let say with $\theta=\theta^\prime$, there will be a corresponding strip in the second half with $\theta=\theta^\prime+\pi/2$. The sum of the shaded area spanned by these strips are given by $\Delta \theta \left(\Delta X_{\theta^\prime}+\Delta X_{\theta^\prime+\pi/2}\right)$. 
\begin{figure}[h]
\centering
\includegraphics[scale=0.55]{figure1.eps}
\caption{Schematic diagram of the projection of standard deviation of the rotated quadrature operator of an arbitrary single-mode quantum state on to the optical tomographic plane. \label{schematic}}
\end{figure}
This process can be repeated for all the horizontal strips in the region between $\theta=0$ to $\theta=\pi$. Two times the sum of the area spanned by all horizontal strips in the region between $\theta=0$ to $\theta=\pi$ gives the total area of the shaded region.

In the limit, $\Delta \theta \rightarrow 0$, the shaded area of the region between $\theta=0$ to $\theta=2\pi$ is calculated to be
\begin{equation}
\int_0^{2\pi} d\theta\, \Delta X_{\theta}=2\times \int_0^{\pi/2} d\theta^\prime\, \left(\Delta X_{\theta^\prime}+\Delta X_{\theta^\prime+\pi/2}\right).\label{shaded}
\end{equation}
The minimum value of the quantity inside bracket in (\ref{shaded}) subjected to the constraint given in equation~(\ref{uncertainty}) is a constant for the pure classical state:
\begin{equation}
\left(\Delta X_{\theta}+\Delta X_{\theta+\pi/2}\right)_{min}=\sqrt{2}.\label{min}
\end{equation}
Therefore, the lower bound of the above integral is evaluated by plugging equation~(\ref{min}) in equation~(\ref{shaded}), and is calculated to be $\sqrt{2}\,\pi$. This lower bound is a constant for the pure single-mode classical state of the field  except for the classical states which are a probabilistic mixture of states. Since the value of the quantity inside bracket in equation~(\ref{shaded}) is always greater than $\sqrt{2}$ for the nonclassical states, the area projected by the standard deviation distribution will  always be  greater than $\sqrt{2}\,\pi$ for all single-mode nonclassical states of the field. Hence, we can use (12) to define a quantity
\begin{equation}
\sigma\left(\ket{\psi}\right)=\int_{0}^{2\pi} d\theta \,\Delta X_\theta,\label{Nonclassical_Area} 
\end{equation}
to characterize the nonclassicality or quantumness of an
arbitrary single-mode quantum state other than the states which are a probabilistic mixture of states. The quantity $\sigma\left(\ket{\psi}\right)$ is the effective area projected by the optical tomogram of the single-mode quantum state $\ket{\psi}$ on the optical tomographic plane.  As mentioned earlier, the effective area $\sigma\left(\ket{\psi}\right)$ has a lower bound for the pure classical state (except for the classical states that are a probabilistic mixture of states), which is $\sqrt{2}\,\pi$. Therefore, the difference $\sigma\left(\ket{\psi}\right)-\sqrt{2}\,\pi$ can be taken as an indicator of nonclassicality 
for the single-mode states of the electromagnetic field. If the projection of the optical tomogram of a pure single-mode quantum state occupies an area more than $\sqrt{2}\,\pi$ on the optical tomographic plane, the state is strictly a nonclassical state. In other words,  a pure single-mode quantum state of light $\ket{\psi}$ is said to be  a nonclassical state if the nonclassical area $\sigma \left(\ket{\psi}\right)-\sqrt{2}\,\pi>0$. 

Let us analyze the effective area $\ket{\psi}$ for the single-mode quantum states of the electromagnetic field. First, consider the case of  a  coherent state $\ket{\alpha}$ (The eigenstate of the annihilation operator, $a\ket{\alpha}=\alpha\ket{\alpha}$, where $\alpha=\modu{\alpha}e^{i\eta}$). The optical tomogram of the coherent state $\ket{\alpha}$ is given by \cite{Lvovsky2009}
\begin{eqnarray}
\omega_{\ket{\alpha}}\left(X_{\theta},\theta\right)=\frac{1}{\sqrt{\pi}} \exp\left[-\left(X_{\theta}-\sqrt{2}\modu{\alpha}\cos(\eta-\theta)\right)^2\right].\nonumber\\\label{tomoCS}
\end{eqnarray}
The projection of optical tomogram $\omega_{\alpha}\left(X_{\theta},\theta\right)$ on the optical tomographic plane displays a structure with single sinusoidal strand \cite{Rohith2016}. The standard deviation in the measurement of the homodyne rotated quadrature operator in the coherent state $\ket{\alpha}$ is calculated to be $\Delta X_\theta=1/\sqrt{2}$. It is straight forward to calculate the effective area spanned by the optical tomogram of coherent state $\ket{\alpha}$ as $\sigma\left(\ket{\alpha}\right)=\sqrt{2}\,\pi$, which is independent of the value of $\alpha$. As $\alpha\rightarrow0$, the coherent state $\ket{\alpha}$ reduces to the vacuum state $\ket{0}$.  
The optical tomogram of the vacuum state $\ket{0}$ is obtained as $\omega_{\ket{0}}\left(X_{\theta},\theta\right)=\exp\left(-X_{\theta}^2\right)/\sqrt{\pi}$, which is a structure with single straight strand in the optical tomographic plane \cite{Rohith2015}. The standard deviation of the rotated quadrature and effective area for the vacuum state $\ket{0}$ is same as that for the coherent state $\ket{\alpha}$. That is, $\Delta X_\theta=1/\sqrt{2}$ and $\sigma\left(\ket{0}\right)=\sqrt{2}\,\pi$. Therefore, the optical tomograms of the pure single-mode classical state of electromagnetic field span a constant effective area of $\sqrt{2}\pi$ on the optical tomographic plane.     As per the derivation of the nonclassical area measure discussed above, a nonzero value of the nonclassical area for a given pure single-mode quantum state is a sufficient condition to say that the state is nonclassical. 
The concept of tomography is based on measuring the observable corresponding to the rotated quadrature operator $X_\theta$ (see the definition given in equation~(\ref{quadratureSingle})). While defining the $X_\theta$, it is already mentioned that  $\theta$   is the phase of the local oscillator in the homodyne setup. A set of quantities corresponding to $X_\theta$ are generated for various phase values of the local oscillator. So, the standard deviation $\Delta X_\theta$ does not correspond to a single quantity of the given field (standard deviation of the quadrature of the field)  alone but to quantities corresponding to the given field and the local oscillator. Hence, the nonclassicality area is beyond squeezing of the field, and we will explicitly show in the remaining part of the paper that this measure also works for nonclassical states that do not show squeezing.

A nonclassicality-inducing operation on the coherent state or the vacuum state yields a nonclassical state for which the nonclassical area spanned by its optical tomogram must be greater than zero. Furthermore, the nonclassical area of a particular nonclassical state must increase with an increase in the strength of the nonclassicality-inducing operations such as the squeezing, photon addition, etc.  By definition, the amount of quantumness associated with a state is independent of the displacement operations and rotations in the phase space. Therefore, the nonclassical area projected by the optical tomogram of the displaced and rotated quantum states must be the same as that of the original state. In the following, we check these key aspects and hence the validity of the nonclassical area measure by calculating the nonclassical area projected by the optical tomograms of different classes of nonclassical states of light.

\subsection{Fock state}
 A simplest state of the optical field to start with would be an $n$-photon Fock state $\ket{n}$, where $n=0,1,\,2,\,\dots,\, \infty$. The optical tomogram of the Fock state $\ket{n}$ is given by 
\begin{equation}
\omega_{\ket{n}}\left(X_{\theta},\theta\right)= Q_{\ket{n}}\left(X_{\theta},\theta\right)Q_{\ket{n}}^\ast\left(X_{\theta},\theta\right), \label{tomo_Fock}
\end{equation}
where $Q_{\ket{n}}\left(X_{\theta},\theta\right)$ is the quadrature representation of the Fock state $\ket{n}$ \cite{Barnett1997}: 
\begin{equation}
Q_{\ket{n}}\left(X_{\theta},\theta\right)=\bra{X_{\theta},\theta}\left.n\right>=\frac{e^{-{X_{\theta}}^2/2}\,H_n\left(X_{\theta}\right)\,e^{-i\, n\theta}}{\pi^{1/4}\,2^{n/2\,\sqrt{n!}}}.
\end{equation}
Here $H_n\left(\cdot\right)$ is the Hermite polynomial of order $n$. The standard deviation in the measurement of the rotated quadrature operator in the $n$-photon Fock state is $\sqrt{n+1/2}$. Therefore, the optical tomogram given in equation~({\ref{tomo_Fock}) span an effective area 
\begin{equation}
\sigma\left(\ket{n}\right)=\sqrt{2(2n+1)}\,\pi, \label{sigma_n}
\end{equation}
on the $X_{\theta}$-$\theta$ plane. The case $n=0$ corresponds to the vacuum state $\ket{0}$, and the equation~(\ref{sigma_n}) once again shows that the nonclassical area $\sigma\left(\ket{0}\right)-\sqrt{2}\,\pi$ spanned by the optical tomogram of the vacuum state $\ket{0}$ is zero. But for the Fock state $\ket{n}$ with $n\neq0$, equation~(\ref{sigma_n})  tells that the degree of nonclassicality associated with the state $\ket{n}$ is always greater than zero, that is $\sigma \left(\ket{n}\right)-\sqrt{2}\,\pi>0$. It also demonstrates that the amount of nonclassicality contained in the state $\ket{n}$ increases monotonically with  the photon number $n$ (see figure~\ref{Nonclassical_fockstate}). Nonclassicality indicator based on the volume of the negative part of the Wigner function of  the Fock state $\ket{n}$ showed that the amount of nonclassicality of the state could be  approximated as $\sqrt{n}/2$ for a large value of $n$ \cite{Kenfack2004}.  The divergence of the amount of nonclassicality of the Fock state $\ket{n}$ with the increase in the photon number $n$ has also been demonstrated using the nonclassicality measures such as entanglement potential \cite{Asboth2005}, the degree of nonclassicality based on characteristic function \cite{Ryl2017}, quantifier based on the quantum superposition principle \cite{Vogel2014}, etc.
\begin{figure}[h]
\centering
\includegraphics[scale=0.4]{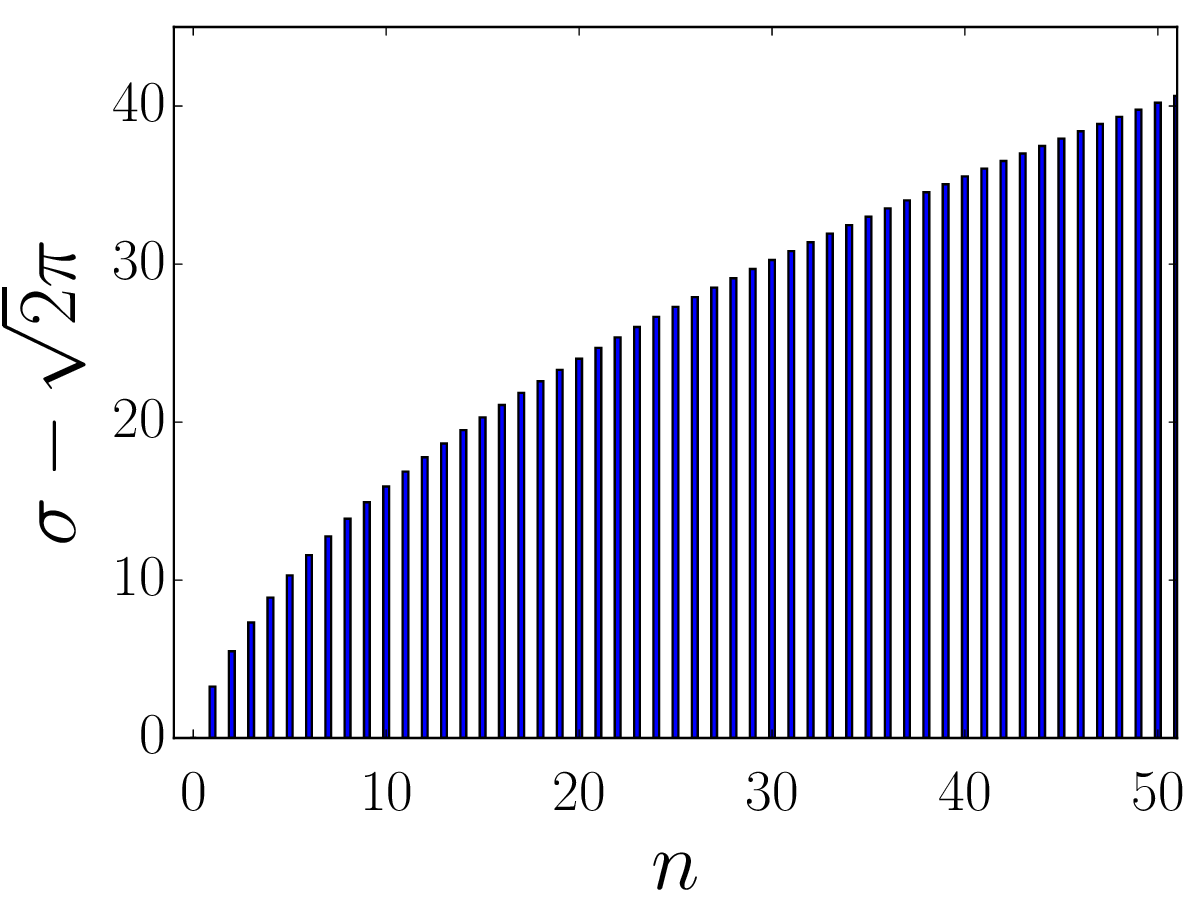}
\caption{Nonclassical area $\sigma-\sqrt{2}\,\pi$ spanned by the optical tomogram of $n$-photon Fock state $\ket{n}$ on the optical tomographic plane as a function of the photon number $n$. Nonclassicality of the state $\ket{n}$ increases monotonically with the increase in the value of $n$. \label{Nonclassical_fockstate}}
\end{figure}

\subsection{Squeezed state}
The action of the squeezing operator $S\left(\xi\right)=\exp\left[\left(\xi^\ast a^2-\xi {a^\dag}^2\right)/2\right]$ on the vacuum state $\ket{0}$ generates a squeezed vacuum state which is defined as
\begin{equation}
\ket{\xi}=S\left(\xi\right)\ket{0},\label{sqzdvac}
\end{equation}
where $\xi=r e^{i\delta}$. Fock state representation of the state $\ket{\xi}$ is given by \cite{Gerry2005}
\begin{equation}
\ket{\xi}=\sum_{n=0}^{\infty} P_{2n}\left(\xi\right)\ket{2n},\label{sq_fock}
\end{equation}
where 
\begin{equation}
P_{2n}\left(\xi\right)=\frac{\left(-1\right)^n}{\sqrt{\cosh r}} \frac{\sqrt{(2n)!}\,e^{in\delta} \left(\tanh r\right)^n}{2^n n!}.
\end{equation}
Substituting equation~(\ref{sq_fock}) in equation~(\ref{opt_tomo_def_purestate}), the optical tomogram of the squeezed vacuum state is obtained as
\begin{eqnarray}
\omega_{\ket{\xi}}\left(X_{\theta},\theta\right)&=&\sum_{n,n^\prime=0}^{\infty}P_{2n}\left(\xi\right)
P_{2n^\prime}^{\ast}\left(\xi\right) Q_{\ket{2n}}\left(X_{\theta},\theta\right)\nonumber\\
 &&\times Q_{\ket{2n^\prime}}^\ast\left(X_{\theta},\theta\right).
\end{eqnarray}
Standard deviation in the measurement of the homodyne quadrature operator  in the squeezed vacuum state $\ket{\xi}$ is calculated as
\begin{equation}
\Delta X_\theta=\sqrt{\left[\cosh (2r)-\sinh (2r) \cos(\delta-2\theta)\right]/2}.\label{Variance_sqvac}
\end{equation}
 Using equations~(\ref{Variance_sqvac}) and (\ref{Nonclassical_Area}), the  effective area spanned by the optical tomogram of the squeezed state on the $X_{\theta}$-$\theta$ plane is evaluated to be
\begin{eqnarray}
\sigma \left(\ket{\xi}\right) =\frac{e^{-r}}{\sqrt{2}}\left[E\left(2\pi-\delta/2\mid k^2\right)+E\left(\delta/2\mid k^2\right)\right],
\end{eqnarray}
where $k^2=\left(1- e^{4r}\right)$ and 
\begin{equation}
E\left(\phi\mid k^2\right)=\int_0^{\phi} \sqrt{1-k^2 \sin^2 \lambda}\, d\lambda,
\end{equation}
is the incomplete elliptical integral of the second kind. The solid line in figure~\ref{fig_sq_vac} shows the nonclassical area $\sigma\left(\ket{\xi}\right)-\sqrt{2}\,\pi$ spanned by the optical tomogram of squeezed vacuum state $\ket{\xi}$ on the optical tomographic plane as a  function of the squeezing parameter $r$. The degree of nonclassicality of the state $\ket{\xi}$ increases monotonically as a function of $r$ and is found to be independent of the value of phase $\delta$. In the limit $r\rightarrow0$, the nonclassical area of the state $\ket{\xi}$ becomes zero which is the value corresponding to the vacuum state $\ket{0}$.
\begin{figure}[h]
\centering
\includegraphics[scale=0.4]{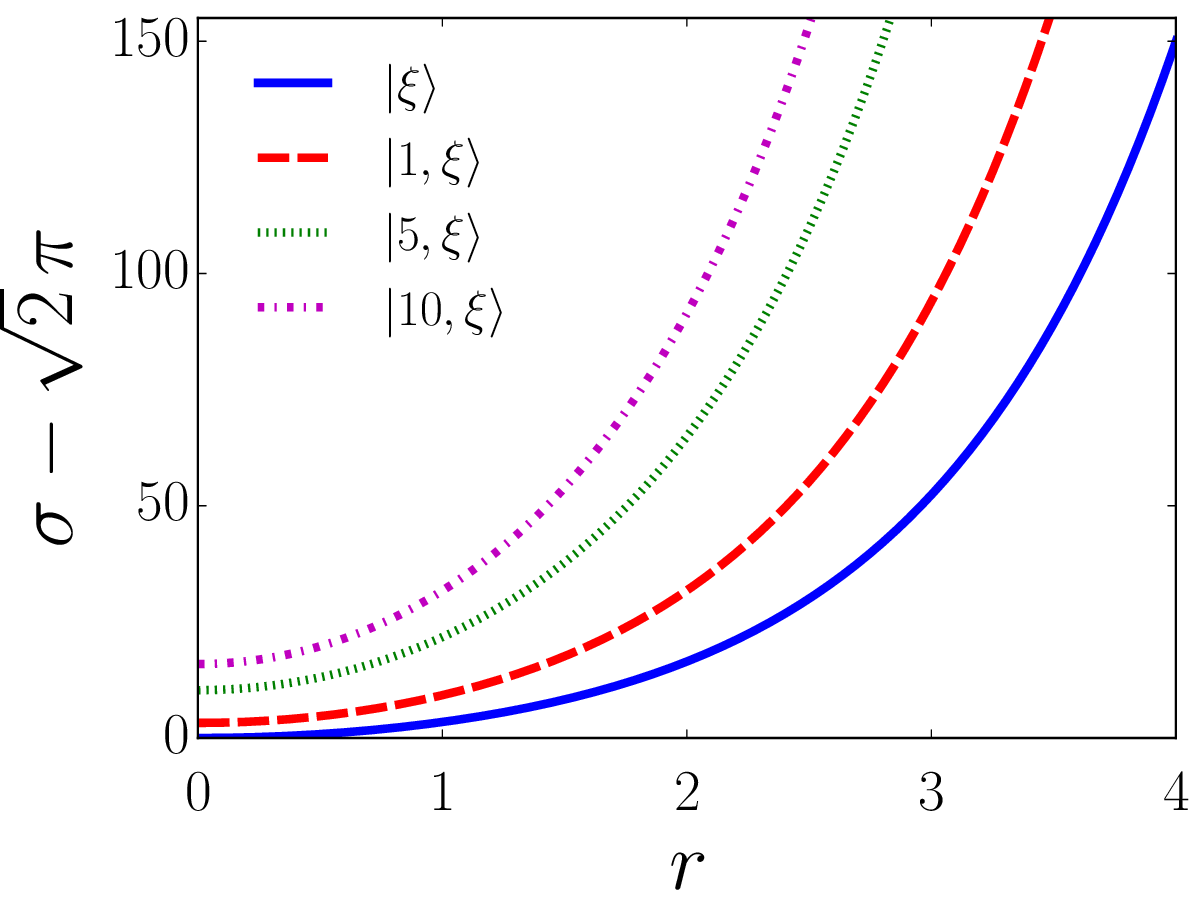}
\caption{Nonclassical area $\sigma-\sqrt{2}\,\pi$ spanned by the optical tomogram of squeezed vacuum state $\ket{\xi}$ (solid line) and squeezed Fock state $\ket{n, \xi}$ with $n=1$ (dashed line), $5$ (dotted line), and $10$ (dash-dot line) as a function of the parameter $r$.  Here $r$ is the argument of the squeezing parameter, $\xi=r\,e^{i\delta}$. The degree of nonclassicality associated with the states $\ket{\xi}$ and $\ket{n,\xi}$ increases monotonically with the  value of $r$ and are independent of the value of $\delta$. In the limit $r\rightarrow0$, the nonclassical area of the state $\ket{n,\xi}$ gives the value corresponding to the Fock state $\ket{n}$. \label{fig_sq_vac}}
\end{figure}

When squeezing operator $S\left(\xi\right)$ acts on the $n$-photon Fock state, it generates the squeezed Fock state  $\ket{n,\xi}$: 
\begin{equation}
\ket{n,\xi}=S\left(\xi\right)\ket{n}=\sum_{j=0}^{\infty} C_j \ket{j}.\label{squeezedFockrepresentation}
\end{equation}
Here the Fock state expansion coefficients are given by \cite{Kral1990}
\begin{eqnarray}
C_j&=&\sqrt{\left(\frac{j!}{n! \mu}\right) \left(\frac{\nu}{2\mu}\right)^{j}}\,\sum_{i=0}^{min\left(n,j\right)} {{n}\choose{i}} \frac{\left(2/\mu\nu\right)^{i/2}}{\left(j-i\right)!} \nonumber\\
&&\times\left(-\frac{\nu^\ast}{2\mu}\right)^{\left(n-i\right)/2}  H_{j-i}\left(0\right)H_{n-i}\left(0\right)
\end{eqnarray}
where $\mu=\cosh(r)$, and $\nu=e^{i\delta} \sinh (r)$. For $n=0$, the state given in equation~(\ref{squeezedFockrepresentation}) retrieves the squeezed vacuum state $\ket{\xi}$. Using ({\ref{squeezedFockrepresentation}), the optical tomogram of the squeezed Fock state  $\ket{n,\xi}$ is calculated as
\begin{equation}
\omega_{\ket{n,\xi}}=\sum_{j,j^\prime=0}^{\infty} C_j C_{j^\prime}^\ast \, Q_{\ket{j}}\left(X_{\theta},\theta\right) Q_{\ket{j}}^\ast\left(X_{\theta},\theta\right).
\end{equation}
For the state $\ket{n,\xi}$, the standard deviation in the measurement of the homodyne quadrature operator is obtained as
\begin{eqnarray}
\Delta X_\theta &=&\sqrt{(n+1/2)}\nonumber\\
&&\times\sqrt{\left[\cosh (2r) -\sinh(2r)\cos(\delta-2\theta)\right]}.\label{std_sqfock}
\end{eqnarray}
Substituting equation~(\ref{std_sqfock}) in equation~(\ref{Nonclassical_Area}), the effective area spanned by the optical tomogram of the squeezed Fock state  $\ket{n,\xi}$ is found to be 
\begin{eqnarray}
\sigma \left(\ket{n,\xi}\right)&=&\sqrt{(n+1/2)}\, e^{-r}\nonumber\\
&&\times\left[E\left(2\pi-\delta/2\mid k^2\right)+ E\left(\delta/2\mid k^2\right)\right].
\end{eqnarray}
It is the product of the effective area spanned by the optical tomogram of the Fock state $\ket{n}$ and that corresponds to a squeezed vacuum state $\ket{\xi}$, scaled down by a factor of $\sqrt{2}\,\pi$. That is, 
\begin{eqnarray}
\sigma \left(\ket{n,\xi}\right) =\left[\sigma \left(\ket{n}\right)\times\sigma \left(\ket{\xi}\right)\right]/\sqrt{2}\,\pi.
\end{eqnarray}
The degree of nonclassicality of the state $\ket{n,\xi}$ then becomes $\sigma \left(\ket{n,\xi}\right)-\sqrt{2}\,\pi$. Variation of the nonclassical area spanned by the optical tomogram of the squeezed Fock states $\ket{n,\xi}$ with $n=1$ (dashed line), $5$ (dotted line), $10$ (dash-dot line),  are shown in figure~\ref{fig_sq_vac}. In general, the nonclassical area of the squeezed Fock state $\ket{n,\xi}$ increases monotonically  the parameter $r$ for all values of $n$. Figure \ref{fig_sq_vac} also shows that, in the limit $r\rightarrow0$, the nonclassical area of the state $\ket{n,\xi}$ gives the value corresponding to the Fock state $\ket{n}$. It is worth noting that the nonclassicality of the state is independent of the value of  $\delta$.

The combined action of the displacement operator $D(\alpha)=\exp\left[\left(\alpha^\ast a-\alpha {a^\dag}\right)/2\right]$ followed by $S\left(\xi\right)$ on the vacuum state generates the squeezed coherent state $\ket{\alpha,\xi}$, for which the amount of nonclassicality of the state is  same as that for the squeezed state $\ket{\xi}$. This feature is essentially captured in the nonclassical area measure as the effective area spanned by the optical tomogram of the  state $\ket{\alpha,\xi}$ on the optical tomographic plane is calculated to be $\sigma \left(\ket{\alpha,\xi}\right) =\sigma \left(\ket{\xi}\right)$. It supports  that the displacement operation does not induce any additional nonclassicality in the squeezed state $\ket{\xi}$. Furthermore, one can show that, $\sigma\left(\ket{\alpha,n,\xi}\right) =\sigma \left(\ket{n,\xi}\right)$. Hence the degree of nonclassicality of the displaced squeezed Fock state $\ket{\alpha,n,\xi}=D(\alpha)S\left(\xi\right)\ket{n}$, is same as that of the squeezed Fock state $\ket{n,\xi}$. Therefore, phase space rotations and displacements of the state will not increase the value of nonclassical area. 

\subsection{Photon-added coherent states}
The addition of photons to a coherent state $\ket{\alpha}$ induces nonclassicality in the resultant states. The nonclassical state generated by the addition of $m$ photons to the coherent field $\ket{\alpha}$ is  called an $m$-photon-added coherent state $\ket{\alpha,m}$, defined as \cite{Agarwal1991,Zavatta2004}
\begin{equation}
\ket{\alpha,m}=N_{\alpha,m}\, {a^{\dag}}^m \ket{\alpha},
\end{equation}
where $N_{\alpha,m}$ is the normalization constant and $\modu{\alpha}^2$ is the mean number of photons in the coherent state $ \ket{\alpha}$. 
The optical tomogram of the  $m$-photon-added coherent states $\ket{\alpha,m}$ obtained by Substituting the Fock state representation of $\ket{\alpha,m}$ in equation~(\ref{opt_tomo_def_purestate}) \cite{Rohith2015}: 
\begin{eqnarray}
\omega_{\ket{\alpha,m}}\left(X_{\theta},\theta\right)&=&\frac{e^{-\modu{\alpha}^2}}{m!L_m(-\modu{\alpha}^2)}\nonumber\\
&&\times \modu{\sum_{n=m}^{\infty}\frac{\alpha^{n-m}\sqrt{n!}\, Q_n\left(X_{\theta},\theta\right)}{(n-m)!}}^2.
\label{optPACS}
\end{eqnarray}
 Here $L_m(x)$ represents the Laguerre polynomial of order $m$. Nonclassical area $\sigma\left(\ket{\alpha,m}\right)-\sqrt{2}\,\pi$ projected by the optical tomogram of $m$-photon-added coherent states on the optical tomographic plane was numerically evaluated using equation~(\ref{Nonclassical_Area}). Figure \ref{Nonclassical_photon-added} shows the variation of the nonclassical area corresponding to the photon-added coherent state as a function of the number of photons $m$ added to the coherent state $\ket{\alpha}$. It clearly shows that the nonclassical area of the state $\ket{\alpha,m}$ increases with the increase in the number of photons added to the coherent field $\ket{\alpha}$.
\begin{figure}[h]
\centering
\includegraphics[scale=0.4]{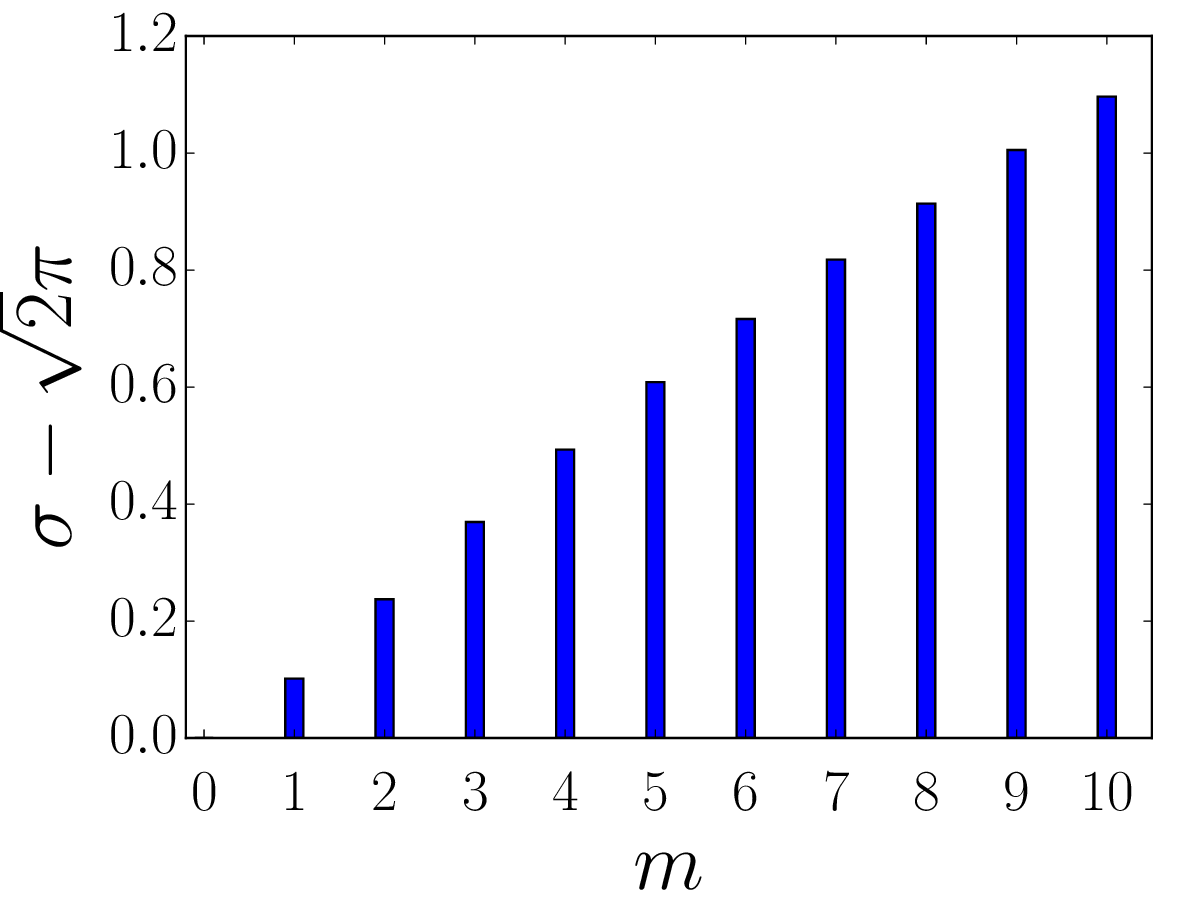}
\caption{Nonclassical area $\sigma-\sqrt{2}\,\pi$ projected by the optical tomogram of $m$-photon-added coherent states $\ket{\alpha,m}$ with $\modu{\alpha}^2=5$ on the optical tomographic plane. The nonclassicality of the state $\ket{\alpha,m}$ increases with the increase in the number of photons added to the coherent field $\ket{\alpha}$.\label{Nonclassical_photon-added}}
\end{figure}

Variation of the nonclassical area projected by the optical tomogram of the state $\ket{\alpha,m}$ as a function of the field intensity $\modu{\alpha}^2$ is shown in figure~\ref{Nuvariation_photon-added}. Here the plots are generated for the cases: $m=1$ (solid line), $5$ (dashed line), and $10$ (dotted line). It is evident from the figure that, for a fixed number of photons $m$, the nonclassicality of the state  $\ket{\alpha,m}$ decreases with an increase in $\modu{\alpha}^2$. For small values of $m$, the state $\ket{\alpha,m}$ becomes more and more coherent with increased  field intensity. This effect is clearly manifested in  figure~\ref{Nuvariation_photon-added}, where it can be seen that the nonclassical area of the state $\ket{\alpha,1}$ approaches to zero (value corresponding to coherent state $\ket{\alpha}$) for $\modu{\alpha}^2$ above $5$. In the limit  $\modu{\alpha}^2\rightarrow\,0$, the nonclassical area gives the value corresponding to   the Fock state $\ket{m}$.
\begin{figure}[h]
\centering
\includegraphics[scale=0.4]{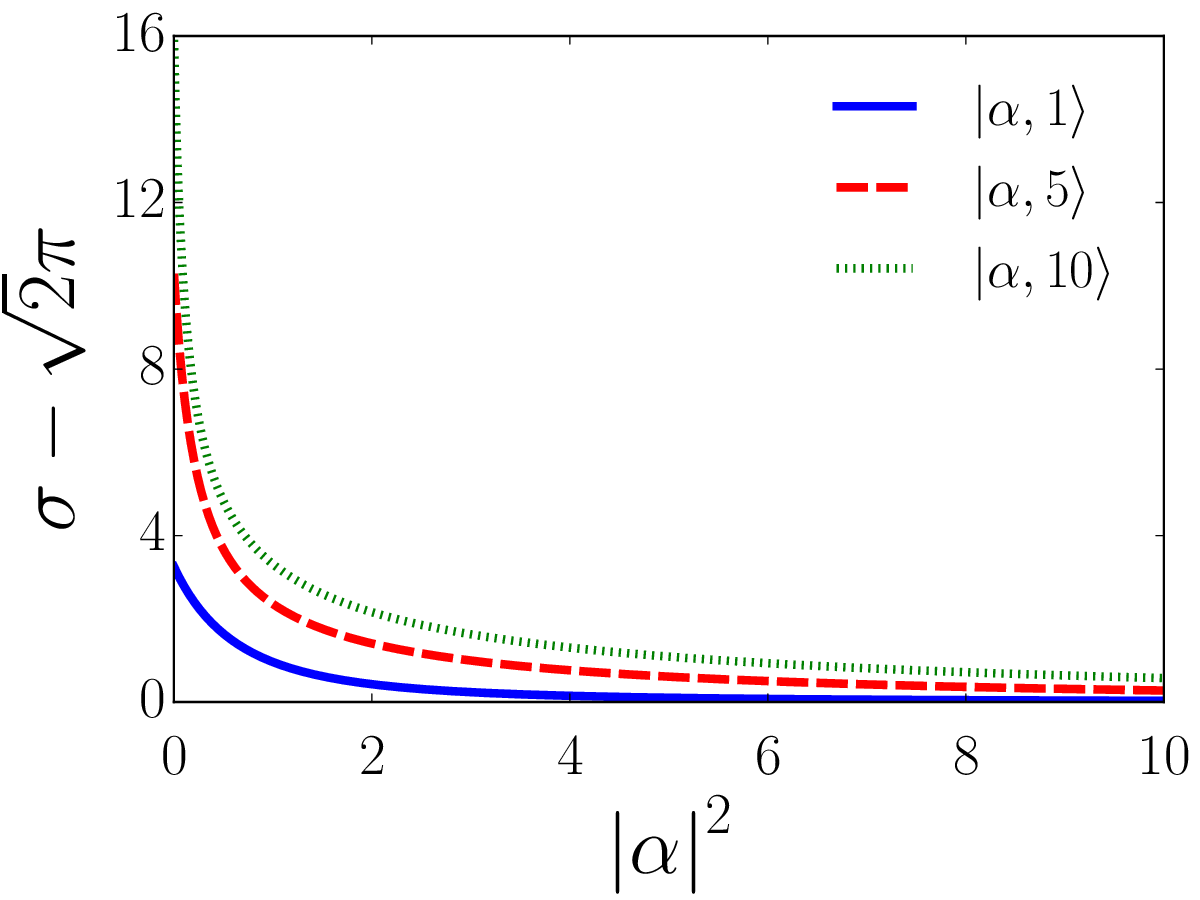}
\caption{Variation of nonclassical area $\sigma-\sqrt{2}\,\pi$ projected by the optical tomogram of $m$-photon-added coherent states $\ket{\alpha,m}$ with respect to the field intensity $\modu{\alpha}^2$ for $m=1$ (solid line), $5$ (dashed line), and $10$ (dotted line). For a fixed number of photons $m$, the nonclasssical area decreases with the increase in the value of $\modu{\alpha}^2$. In the limit $\modu{\alpha}^2\rightarrow\,0$, nonclassical area gives the value corresponding to the that for the Fock state $\ket{m}$. \label{Nuvariation_photon-added}}
\end{figure}

\subsection{Even and odd coherent states}
Nonclassical states can also be obtained by superimposing two or more classical states. Consider the nonclassical states generated by the superposition of two coherent states with $\pi$ phase difference between them. Specifically, we consider the states defined by:
\begin{equation}
\ket{\alpha}_{h}=N_{h}\left[\ket{\alpha}+e^{i\pi h}\ket{-\alpha}\right],\label{EvenOdd}
\end{equation}
 where $N_h$ is the appropriate normalization constant, and the state with $h=0$ corresponds to the even coherent state, and $h=1$ corresponds to the odd coherent state. A suitable transformation in the symplectic tomogram of the even and odd coherent states \cite{Mancini1996} gives the corresponding optical tomogram of the states. The optical tomogram corresponds to the state $\ket{\alpha}_{h}$ is given by \cite{Rohith2015}
\begin{eqnarray}
\omega_{\ket{\alpha}_h}\left(X_{\theta},\theta\right)&=&N_{h}^2 \modu{\sum_{r=0}^{1} e^{i\pi r h} Q_{\ket{\alpha e^{i\pi r}}}\left(X_{\theta},\theta\right)}^2,\label{Opt_EOCS}
\end{eqnarray}
where
\begin{eqnarray}
Q_{\ket{\alpha e^{i\pi r}}}\left(X_{\theta},\theta\right)&=&\frac{1}{\pi^{1/4}}\exp\left[-\frac{X_{\theta}^2}{2}-\frac{(\alpha e^{i\pi r})^2\,e^{-i\,2\theta}}{2}\right.\nonumber\\
&&\left.-\frac{\modu{\alpha}^2}{2}+\sqrt{2}\,\alpha e^{i\pi r}\, X_{\theta} e^{-i\,\theta}\right]\label{quadrature_coherent}
\end{eqnarray}
 is the quadrature representations of the coherent states $\ket{\alpha e^{i\pi r}}$ \cite{Barnett1997}.  We have numerically computed the nonclassical area spanned by the optical tomogram given in equation~(\ref{Opt_EOCS}) on the $X_{\theta}$-$\theta$ plane as a function of $\modu{\alpha}^2$ (see figure~\ref{fig_Opt_EOCS}). It is noted that the nonclassical area (degree of nonclassicality) of the state $\ket{\alpha}_{h}$ is independent of the value of the argument of $\alpha$. For large $\modu{\alpha}^2$ values, the nonclassical area associated with the even and odd coherent states is the same, consistent with the results in \cite{VanEnk2003,Rohith2016}. It has been shown that for large $\modu{\alpha}^2$ values, the coherent states appearing in the superposition can be taken as orthonormal basis for representing the two-mode entangled states generated by the action of a $50/50$ beam splitter on the state $\ket{\alpha}_{h}$ with vacuum state taken in the other input arm \cite{VanEnk2003}. Here the amount  entanglement of the states generated at the output of the beam splitter is also a measure of the amount of nonclassicality of the input state. For large $\modu{\alpha}^2$ values, the entangled states generated by the even and odd coherent state have the same amount of entanglement \cite{Rohith2016}. The nonclassical area measure essentially captures this feature-note that the  two curves in figure~\ref{fig_Opt_EOCS} merge for $\modu{\alpha}^2$ values greater than $3$. 
\begin{figure}[h]
\centering
\includegraphics[scale=0.4]{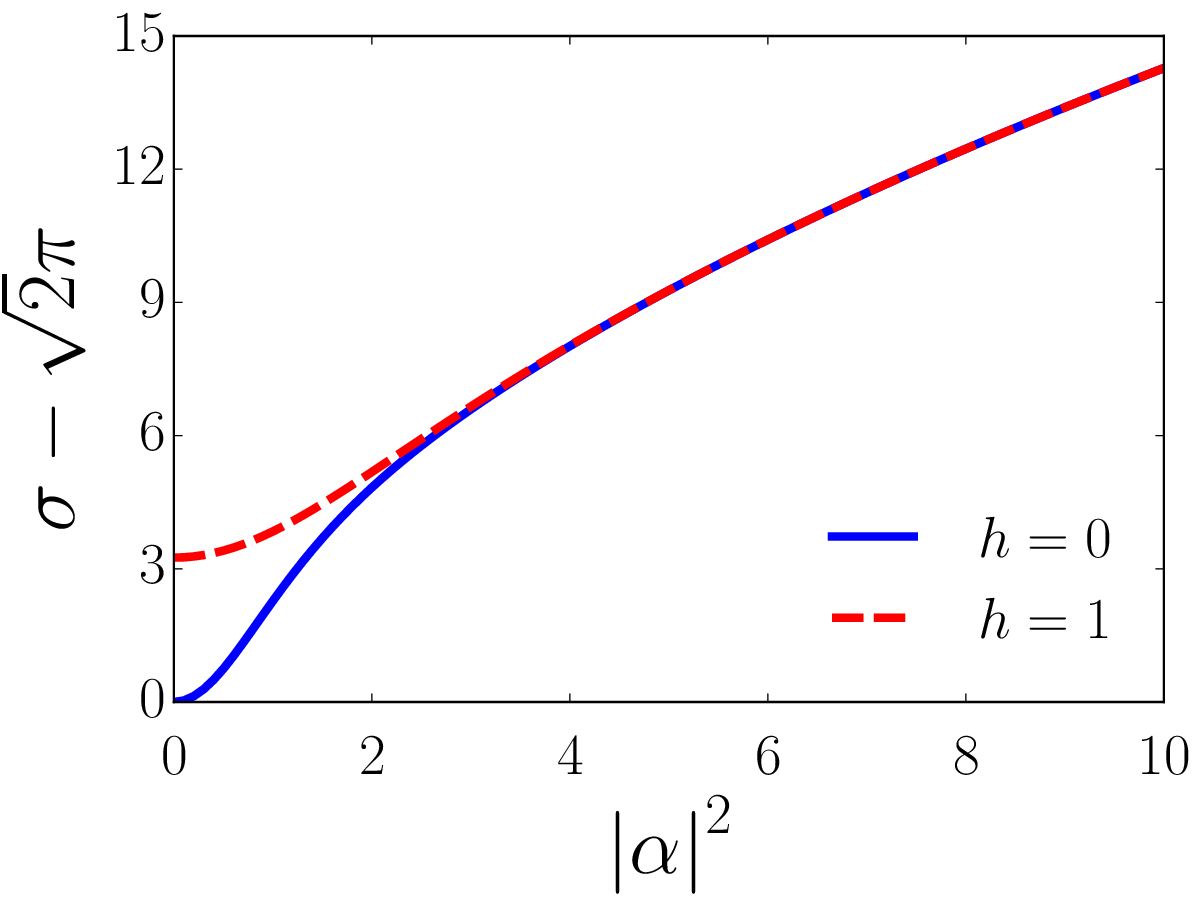}
\caption{Nonclassical area $\sigma-\sqrt{2}\,\pi$ spanned by the optical tomogram of the even (solid line) and odd (dashed line) coherent states $\ket{\alpha}_h$ on the optical tomographic plane as function of $\modu{\alpha}^2$. For large $\modu{\alpha}^2$ values, the degree of nonclassicality associated with the even and odd coherent states are the same. \label{fig_Opt_EOCS}}
\end{figure}

In this section, we have defined nonclassical area for a single-mode field and described the calculations of the  nonclassicality indicator for a single-mode field using the  nonclassical area.  Next, we investigate the effect of decoherence of the single-mode state of the field on the nonclassical area.

\subsection{Effect of decoherence on the nonclassical area}
Here we check the robustness of the nonclassical area indicator of nonclassicality by investigating the effect of environment-induced decoherence of the single-mode field on the nonclassical area measure. We consider an amplitude decay model of decoherence of the state due to the interaction of the field mode with the external environment. We model the external environment as a collection of infinite  harmonic oscillators maintained at zero temperature. Within Born-Markov approximation, the zero temperature master equation in the interaction picture can be written as \cite{Gardiner1991}
\begin{equation}
\frac{\partial\rho}{\partial t}=\gamma\left(2\,a\rho a^\dag-a a^\dag \rho-\rho a^\dag a\right),\label{master}
\end{equation}
where $\gamma$ is the interaction strength of the field mode with the external environment. It should be noted that the equation~(\ref{master}) only considers the
decoherence induced by the coupling to the quantum vacuum and the thermal-induced decoherence is excluded. The above master equation can be solved using the Laplace transform method and the density matrix at time $t$ can be calculated in the Fock basis as
\begin{equation}
\rho\left(t\right)=\sum_{n,\,n^\prime} \rho_{n n^\prime}\ket{n}\bra{n^\prime}, \label{rho_dec}
\end{equation}
where the Fock state expansion coefficients are given by \cite{Biswas2007}
\begin{eqnarray}
\rho_{n, n^\prime}&=&e^{-\gamma t (n+n^\prime)}\sum_{r=0}^{\infty}\sqrt{\left(^{n+r}C_{r}\right)\, \left(^{n^\prime+r}C_{r}\right)} \left(1-e^{-2\gamma t}\right)^r\nonumber\\
&&\times \rho_{n+r, n^\prime+r} \left(t=0\right).
\end{eqnarray}
We have numerically computed the optical tomogram [using (\ref{opt_tomo_def})], and hence the nonclassical area spanned by the optical tomogram of state given in equation~(\ref{rho_dec}) for various initial states. Figure \ref{decoherenceSingleMode} shows the exponential decay of nonclassical area for various initial states due the interaction of the system with the external environment.
\begin{figure}[h]
\centering
\includegraphics[scale=0.53]{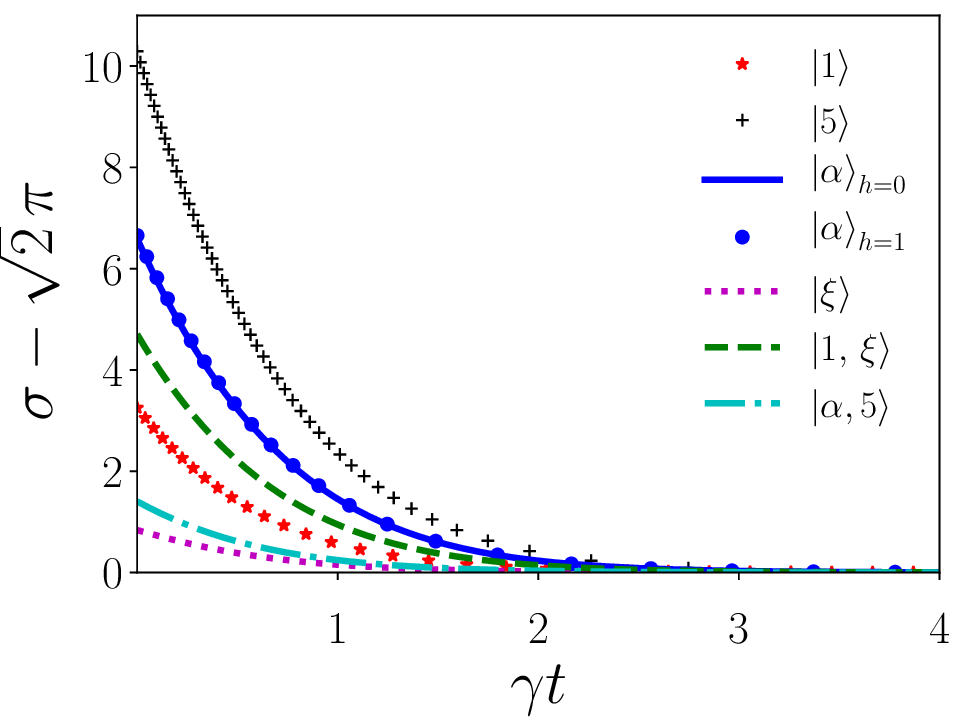}
\caption{Exponential decay of nonclassical area corresponding to $1$-photon Fock state $\ket{1}$ (red marker), $5$-photon Fock state ${\ket{5}}$ (plus marker), even ($h=0$, blue solid-line) and odd ($h=1$, blue circle marker) coherent state $\ket{\alpha}_h$ with $\modu{\alpha}^2=3$, squeezed vacuum state $\ket{\xi}$ (magenta dotted-line) and $1$-photon-added squeezed state (green dashed-line) with $\xi=0.5 i$, and $5$-photon-added coherent state $\ket{\alpha,5}$ with $\modu{\alpha}^2=2$ (cyan dash-dot line), as a function of time $\gamma t$.}\label{decoherenceSingleMode}
\end{figure}
As seen from figure~\ref{decoherenceSingleMode}, the solution  of equation~(\ref{master}) shows
that for all input states without distinction the nonclassical area
exponentially decays with characteristic time $1/\gamma$ which indicates that the nonclassical area indicator of nonclassicality is robust against environment-induced decoherence.

\section{Nonclassical area for a two-mode field }\label{sec4}

In this section, we extend our analysis of the  nonclassical area  given in section~\ref{sec3} and introduce an indicator of nonclassicality for a two-mode state of electromagnetic field. 
The optical tomogram of several two-mode states was theoretically calculated using equation~(\ref{opt_tomo_def_multimode}) \cite{Rohith2016,Sharmila2019}. Since the homodyne rotated quadrature operator $\mathbb{X}_{\theta_1,\theta_2}$ [see (\ref{quadratureMultimode})] is in a separable form, and it obeys the uncertainty relation, it is straightforward to extend our analysis using the uncertainty relation for the case of a two-mode field and arrive at the following expression for the effective area projected by the optical tomogram of a generic two-mode state $\ket{\Psi}$:
\begin{equation}
\sigma\left(\ket{\Psi}\right)=\int_{0}^{2\pi} \,\int_{0}^{2\pi}\,d\theta_1 \,d\theta_2 \,\Delta X_{\theta_1,\,\theta_2}, \label{Nonclassical_Area_twomode} 
\end{equation}
where $\Delta X_{\theta_1,\,\theta_2}$ is the standard deviation in the measurement of the operator $\mathbb{X}_{\theta_1,\theta_2}$. As mentioned earlier, the effective area $\sigma\left(\ket{\Psi}\right)$ has a lower bound for the pure classical states such as a two-mode coherent state $\ket{\alpha_1}\otimes\ket{\alpha_2}$ ($\alpha_1,\,\alpha_2 \in \mathbb{C}$) or vacuum state $\ket{0}\otimes\ket{0}$. The effective area spanned by the optical tomogram of both of these states are calculated to be $2\sqrt{2}\, \pi^2$. All the two-mode nonclassical states will be having an effective area greater than $2\sqrt{2}\, \pi^2$. Therefore, the deviation
\begin{equation}
\sigma\left(\ket{\Psi}\right)-2\sqrt{2}\, \pi^2,
\end{equation}
 can be taken as the nonclassical area indicator of nonclassicality associated with an arbitrary two-mode state other than the states which are a probabilistic mixture of two-mode states. A non-zero value of the nonclassical area $\sigma\left(\ket{\Psi}\right)-2\sqrt{2}\, \pi^2$ is a sufficient condition for an arbitrary pure two-mode state of the electromagnetic field to be nonclassical. The amount of nonclassicality associated with a  state is proportional to the nonclassical area spanned by the optical tomogram of the corresponding state on the optical tomographic plane. Next, we calculate the nonclassical area for specific well-known two-mode fields such as a two-mode squeezed vacuum state and two-mode even and odd coherent states and see whether the nonclassicality of these states is reflected in it. We also check the robustness of the nonclassical area measure  by investigating the effect of decoherence of the two-mode state on the nonclassical area.

\subsection{Two-mode squeezed vacuum state}
The action of a two-mode squeeze operator  $S_2\left(\xi\right)=\exp\left(\xi^\ast a_1\, a_2-\xi {a_1^\dag a_2^\dag}\right)$, where $a_1$ and $a_2$ are the operators for the two modes and $\xi=r\,e^{i\delta}$, on the two-mode vacuum state $\ket{0}_{a_1}\ket{0}_{a_2}=\ket{0,0}$ generates a two-mode squeezed vacuum state $\ket{\xi}_2$. In the Fock basis representation the two-mode squeezed vacuum state can be written as \cite{Gerry2005}:
\begin{eqnarray}
\ket{\xi}_2=\frac{1}{\cosh r}\sum_{n=0}^{\infty} \left(-1\right)^n e^{i\,n\delta} \left(\tanh r\right)^n \ket{n,n}. \label{twomode-squeezed-state}
\end{eqnarray}
It is known that the state $\ket{\xi}_2$ is highly nonclassical state exhibiting quadrature squeezing and entanglement \cite{Hiroshima2001}. Using equation~(\ref{opt_tomo_def_multimode}), the optical tomogram of the state $\ket{\xi}_2$ is calculated  to be
\begin{eqnarray}
&\omega_{\ket{\xi}_2}&\left(X_{\theta_1},\theta_1; X_{\theta_2},\theta_2\right)=\frac{1}{\cosh^2 r}\sum_{n,\,n^\prime=0}^{\infty} \left(-1\right)^{n+n^\prime} \nonumber\\
&&\times e^{i\,(n-n^\prime)\delta}\left(\tanh r\right)^{n+n^\prime}Q_{\ket{n}}\left(X_{\theta_1},\theta_1\right)\nonumber\\
&&\times  Q_{\ket{n^\prime}}^\ast\left(X_{\theta_1},\theta_1\right) Q_{\ket{n}}\left(X_{\theta_2},\theta_2\right)Q_{\ket{n^\prime}}^\ast\left(X_{\theta_2},\theta_2\right).\label{tomogram-two-SqVac}
\end{eqnarray}
The variance of the homodyne rotated quadrature operator $\mathbb{X}_{\theta_1,\theta_2}$ in the state $\ket{\xi}_2$ is calculated as
\begin{eqnarray}
\left(\Delta X_{\theta_1,\,\theta_2}\right)^2&=& 
\frac{1}{2}\left[2\sinh^2(r)-\sinh(2r) \cos\left(\delta-\theta_{1}-\theta_{2}\right)\right.\nonumber\\
&&\left.+1\right].\label{variance_xi_2}
\end{eqnarray}
We have numerically computed the nonclassical area of the state $\ket{\xi}_2$ using equations~(\ref{variance_xi_2}) and (\ref{Nonclassical_Area_twomode}) and is plotted as a function of the squeezing parameter $r$ in figure~\ref{fig_twomodeSqvacuum}.
\begin{figure}[h]
\centering
\includegraphics[scale=0.5]{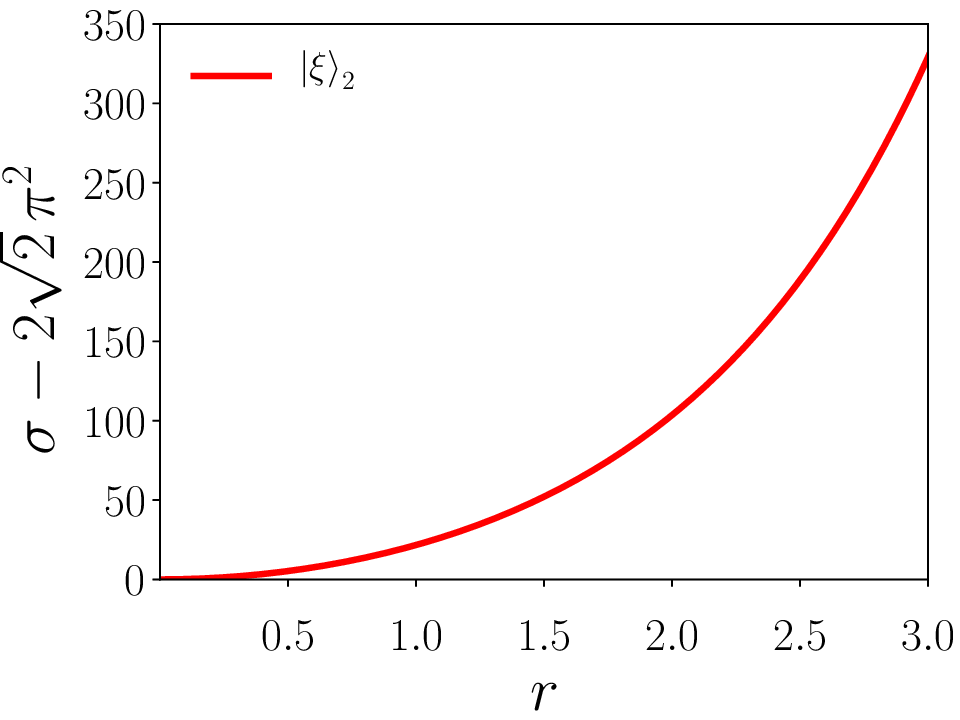}
\caption{Variation of the nonclassical area spanned by the optical tomogram of two-mode squeezed vacuum state $\ket{\xi}_2$ as a function of the squeezing parameter $r$.}\label{fig_twomodeSqvacuum}
\end{figure}
As we know from the other nonclassicality measures, the quantumness of the state $\ket{\xi}_2$ increases with increases in the squeezing parameter $r$ and this trend is well reflected in the nonclassical area indicator of nonclassicality.

\subsection{Two-mode even and odd coherent state}
Next, we look at the two-mode version of the even and odd coherent states given in equation~(\ref{EvenOdd}). A beam splitter arrangement can be used to generate the two-mode even and odd coherent state by taking a single-mode even and odd coherent state at one input port and the vacuum at the other input port \cite{Rohith2016}. The state at the output of the beam splitter is the two-mode even and odd coherent state given by
\begin{equation}\label{EvenOdd_twomode}
\ket{\Psi}_h^{(2)}=N_{h}\left[\ket{\alpha}\ket{\alpha}+(-1)^{h}\ket{-\alpha}\ket{-\alpha}\right],
\end{equation}
where $N_{h}$ is the normalization constant [same as in equation~(\ref{EvenOdd})],  $\alpha=\modu{\alpha}e^{i\eta}$ and $h=0$ (even state) or $1$ (odd state). The two-mode even and odd coherent state is a nonclassical state having entanglement property. In terms of the quadrature representation of a single-mode coherent given in equation~(\ref{quadrature_coherent}), the optical tomogram of the state $\ket{\Psi}_h^{(2)}$ is given by \cite{Rohith2016}
\begin{eqnarray}
&&\omega_{\ket{\Psi}_h^{(2)}}\left(X_{\theta_1},\theta_1; X_{\theta_2},\theta_2\right)\nonumber\\
&&=N_h^2\left|\sum_{r=0}^{1} Q_{\ket{\alpha e^{i\pi r}}}\left(X_{\theta_1},\theta_1\right) Q_{\ket{\alpha e^{i\pi r}}}\left(X_{\theta_2},\theta_2\right)  \right|^2.
\end{eqnarray}
 The variance of the two-mode rotated quadrature $\mathbb{X}_{\theta_1,\theta_2}$ in the state $\ket{\Psi}_h^{(2)}$ is found to be
\begin{eqnarray}\label{var_evenOdd_twomode}
&&\left(\Delta X_{\theta_1,\,\theta_2}\right)^2= \frac{1}{2}+\modu{\alpha}^{2}\left\{\cos^{2}(\eta-\theta_{1})+\cos^{2}(\eta-\theta_{2})\right.\nonumber\\
&&+\left[\tanh(2\modu{\alpha}^{2})\right]^{(-1)^h}\left[1+\cos(\theta_{1}-\theta_{2})\right]\nonumber\\
&&-1+\cos(2\eta-\theta_{1}-\theta_{2})\left.\right\}.
\end{eqnarray}
\begin{figure}[h]
\centering
\includegraphics[scale=0.5]{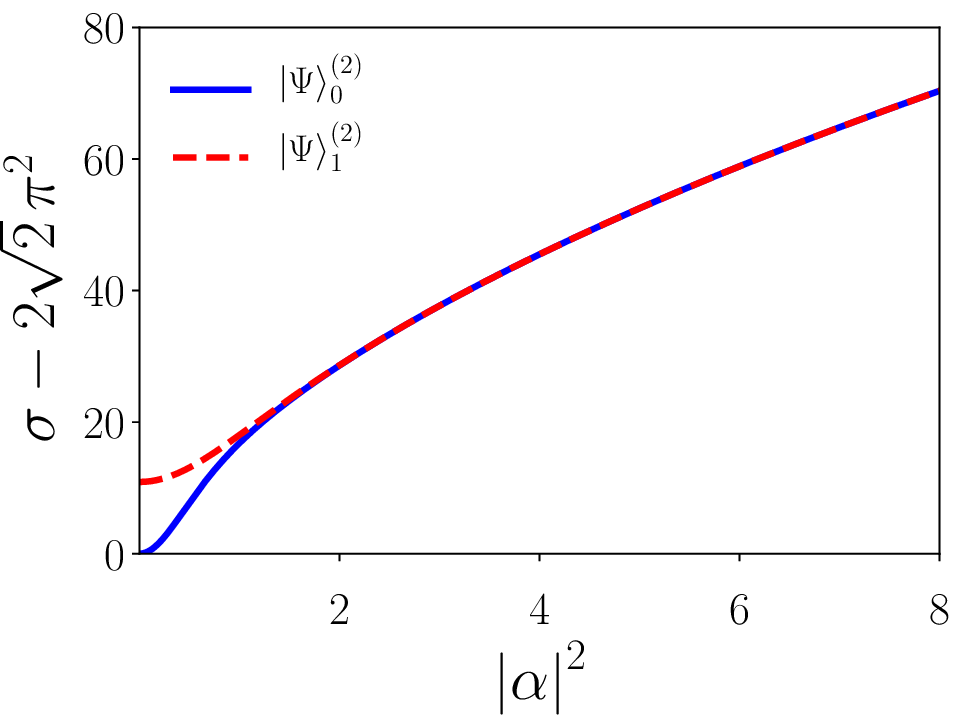}
\caption{Variation of the nonclassical area spanned by the optical tomogram of two-mode even $\ket{\Psi}_0^{(2)}$ and odd $\ket{\Psi}_1^{(2)}$ coherent states as a function of the field strength $\modu{\alpha}^2$.}\label{fig_twomodeEOCS}
\end{figure}
The nonclassical area spanned by the optical tomogram $\omega_{\ket{\Psi}_h^{(2)}}\left(X_{\theta_1},\theta_1; X_{\theta_2},\theta_2\right)$ on the optical tomographic plane was numerically calculated using equations~(\ref{var_evenOdd_twomode}) and (\ref{Nonclassical_Area_twomode}). Figure~\ref{fig_twomodeEOCS} shows the variation of the nonclassical area of the state $\ket{\Psi}_h^{(2)}$ as a function of the field strength $\modu{\alpha}^2$. It is evident that the values of nonclassical area for the states $\ket{\Psi}_0^{(2)}$ and $\ket{\Psi}_1^{(2)}$ are the same for  field strengths above $\modu{\alpha}^2 \approx 1.8$. Using the entanglement measure, it was shown that the nonclassicality of the two-mode even and odd coherent states have the same value for large field strengths. In the limit $\modu{\alpha}^2\rightarrow 0$, the amount of quantumness of the state $\ket{\Psi}_0^{(2)}$ vanishes whereas for $\ket{\Psi}_1^{(2)}$, it is a finite value  \cite{Rohith2016}. These features are nicely captured in the nonclassical area indicator of nonclassicality (see figure~\ref{fig_twomodeEOCS}).  These observations again emphasize that the nonclassical area is a promising candidate to study the nonclassicality of quantum states of light with the added advantage that it can be measured directly from the standard deviation in the measurements of homodyne rotated quadrature operator.

\subsection{Effect of decoherence}
In this section, we study the effect of environment-induced decoherence of the state on the nonclassical area measure of nonclassicality for a two-mode field.  We consider the two-mode extension of the zero-temperature master equation given in equation~(\ref{master}) to model the amplitude decay of a two-mode field. Here also we study the decoherence induced by the coupling of a two-mode field to the quantum vacuum and exclude the thermal-induced decoherence.  The zero-temperature master equation for a two-mode field with density matrix $\rho^{(2)}$ is given by \cite{Gardiner1991,Srinivasan1991}
 \begin{equation}
\frac{\partial\rho^{(2)}}{\partial t}=\sum_{k=1}^{2} \gamma_k\left(2\,a_k\rho^{(2)} a_k^\dag-a_k a_k^\dag \rho^{(2)}-\rho^{(2)} a_k^\dag a_k\right),\label{mastertwomode}
\end{equation}
where $\gamma_k$'s are the interaction strengths of the field mode $a_k$ with the external environment. One can analytically solve the master equation (\ref{mastertwomode}) using the disentagling theorem for SU(1,1) in thermofield dynamics notation \cite{Srinivasan1991}. The density matrix elements of the two-mode field $\rho^{(2)}$ at any instant $t$ can be written in the Fock basis as 
\begin{eqnarray}
&&\bra{m_1,\,m_2} \rho^{(2)}\left(t\right) \ket{n_1,\, n_2}=\sum_{r_1,\,r_2=0}^{\infty} \Gamma_1\, \Gamma_2 \nonumber\\
&&\times\bra{m_1+r_1, m_2+r_2} \rho^{(2)}\left(0\right)\ket{n_1+r_1, n_2+r_2},
\end{eqnarray}
where
\begin{eqnarray}
\Gamma_j&=&\sqrt{^{m_j+r_j}C_{r_j} \,\, ^{n_j+r_j}C_{r_j}}\, \left[1-\exp\left(-2\gamma_j t\right)\right]^{r_j}\nonumber\\
&&\times \exp\left[-\gamma_j t \left(m_j+n_j\right)\right],
\end{eqnarray}
and $\rho^{(2)}\left(0\right)$ is the initial density matrix of the two-mode field. Using the above equation and (\ref{opt_tomo_def_multimode}), we can evaluate the optical tomogram of the state $\rho^{(2)}\left(t\right)$, which is undergoing amplitude decay due to interaction with the external environment. We have numerically calculated the standard deviation in the measurement of rotated quadrature operator $\mathbb{X}_{\theta_1,\theta_2}$ and hence the nonclassical area of the state $\rho^{(2)}\left(t\right)$, which is initially prepared in a two-mode squeezed vacuum state as well as two-mode even and odd coherent states, as a function of time (see figure~\ref{fig_twomodeDec}). The figure shows that at the initial time, the nonclassical area of the state $\rho^{(2)}\left(t\right)$ for initial $\ket{\Psi}_h^{(2)}$ and $\ket{\xi}_2$ give the value corresponding to the pure states $\ket{\Psi}_h^{(2)}$ and $\ket{\xi}_2$, respectively. As the time progress the nonclassical area decays smoothly and finally becomes zero for both the initial states. In the long time limit, both the two two-mode states reduce to the two-mode vacuum state $\ket{0,0}$ for which the  nonclassical area is zero. 
\begin{figure}[h]
\centering
\includegraphics[scale=0.5]{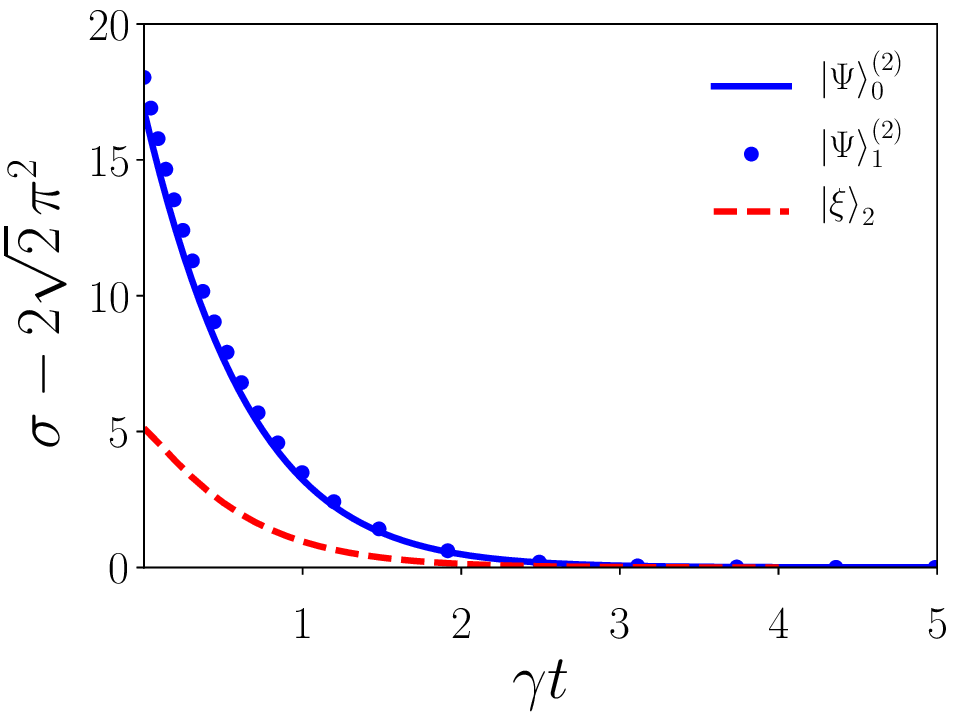}
\caption{Decay of the nonclassical area corresponding to initial two-mode even $\ket{\Psi}_0^{(2)}$ (blue solid line) and odd $\ket{\Psi}_1^{(2)}$ (blue dots) coherent states with $\modu{\alpha}^2=1$, and initial two-mode squeezed state $\ket{\xi}_2$ (red dashed line) with $\xi=0.5 i$, as a function of the interaction time $\gamma t$. }\label{fig_twomodeDec}
\end{figure}

\section{Nonclassical area of $p$-mode states and mixed states}\label{sec5}
Here we generalize the results described in the previous sections to the case of a generic $p$-mode state $\ket{\Psi}^{(p)}$ of the electromagnetic field. Since the $p$-mode rotated quadrature operator given in equation~(\ref{quadratureMultimode}) is in a separable form, our analysis using uncertainty relation given in section~\ref{sec3} can be repeated in a straightforward manner to arrive at an expression for the effective area projected by the optical tomogram [given in equation~(\ref{opt_tomo_def_multimode})] of a $p$-mode field  as
\begin{equation}
\sigma\left(\ket{\Psi}^{(p)}\right)=\int_{0}^{2\pi} d\theta_1\, d\theta_2\,  d\theta_3\, \cdots\, d\theta_p \, \Delta X_{\theta_1,\,\theta_2,\, \theta_3,\, \cdots,\, \theta_p},\label{Nonclassical_AreaPmode} 
\end{equation} 
where $\Delta X_{\theta_1,\,\theta_2,\, \theta_3,\, \cdots,\, \theta_p}$ is the standard deviation in the measurement of $p$-mode rotated quadrature operator $\mathbb{X}_{\theta_1,\,\theta_2,\,  \dots,\, \theta_p}$ in the state $\ket{\Psi}^{(p)}$.
The effective area $\sigma\left(\ket{\Psi}^{(p)}\right)$ has a lower bound for the pure $p$-mode classical state of the field ($p$-mode coherent state) which is calculated to be $2^{\left(p-1/2\right)}\pi^p$. All the $p$-mode nonclassical states, other than the states which are a probabilistic mixture of $p$-mode states, will have an effective area which is greater than $2^{\left(p-1/2\right)}\pi^p$. Therefore, the nonclassical  area indicator of nonclassicality for a $p$-mode state can be written as: 
\begin{equation}
\sigma \left(\ket{\ket{\Psi}^{(p)}}\right)-2^{\left(p-1/2\right)}\pi^p.\label{p-modeNA}
\end{equation}
A nonzero value of nonclassical area $\sigma \left(\ket{\ket{\Psi}^{(p)}}\right)-2^{\left(p-1/2\right)}\pi^p$ for a pure $p$-mode field is a sufficient condition to say that state is nonclassical. For a $p$-mode coherent state or a $p$-mode vacuum state, which are considered to be the classical states of the field, the nonclassical area is zero. 

 When the state is a probabilistic mixture of states, the standard deviation in the measurement of rotated quadrature operator, $\Delta X_{\theta_1,\,\theta_2,\, \theta_3,\, \cdots,\, \theta_p}$ is a hybrid quantity which contains classical mixing as well as quantum uncertainty in it. Though the probabilistic mixture of states such as the thermal states, mixture of coherent states, etc, are considered to be classical states, the classical mixing contribution to $\Delta X_{\theta_1,\,\theta_2,\, \theta_3,\, \cdots,\, \theta_p}$ in such states may lead to a nonzero value for the nonclassical area. If one could remove the classical mixing contribution from the standard deviation $\Delta X_{\theta_1,\,\theta_2,\, \theta_3,\, \cdots,\, \theta_p}$, it is possible to follow the procedure described in this section to find the degree of nonclassicality of the states which are probabilistic mixture of states. One must replace the standard deviation $\Delta X_{\theta_1,\,\theta_2,\, \theta_3,\, \cdots,\, \theta_p}$ in equation~(\ref{Nonclassical_AreaPmode}) with a modified standard deviation which does not have the classical mixing contribution in it. For example, in the case of a thermal state, one could use the uncertainty relation for mixed states \cite{Luo2005}, introduced using the Wigner-Yanase skew information,  to perform the calculation of nonclassical area. In this approach, the skew information for the measurement of the observable $X_{\theta_1,\,\theta_2,\, \theta_3,\, \cdots,\, \theta_p}$ in the state $\rho$, given by
	\begin{equation}
	I=-\frac{1}{2}{\rm Tr}\left[\sqrt{\rho}\,X_{\theta_1,\,\theta_2,\, \theta_3,\, \cdots,\, \theta_p}\right]^2,
	\end{equation}
is used to modify the variance in the measurement of $X_{\theta_1,\,\theta_2,\, \theta_3,\, \cdots,\, \theta_p}$ as
\begin{eqnarray}
&&\bar{\Delta}^2 X_{\theta_1,\,\theta_2,\, \theta_3,\, \cdots,\, \theta_p}\nonumber\\
&&=\sqrt{\Delta^4 X_{\theta_1,\,\theta_2,\, \theta_3,\, \cdots,\, \theta_p}-\left[\Delta^2 X_{\theta_1,\,\theta_2,\, \theta_3,\, \cdots,\, \theta_p}-I\right]^2}.
\end{eqnarray}
The uncertainty relation now becomes \cite{Luo2005}
\begin{equation}
\bar{\Delta}X_{\theta_1,\,\theta_2,\, \theta_3,\, \cdots,\, \theta_p}\bar{\Delta}X_{\theta_1+\pi/2,\,\theta_2+\pi/2,\, \theta_3+\pi/2,\, \cdots,\, \theta_p+\pi/2}\geq\frac{1}{2}.
\end{equation}
In the following we show that, by replacing $\Delta X_{\theta_1,\,\theta_2,\, \theta_3,\, \cdots,\, \theta_p}$ in equation~(\ref{Nonclassical_AreaPmode}) with the modified standard deviation $\bar{\Delta}X_{\theta_1,\,\theta_2,\, \theta_3,\, \cdots,\, \theta_p}$, the nonclassical area for a thermal state is zero. 

For simplicity, we consider a single-mode thermal state with a mean photon number $\bar{n}$. It can be expressed in the Fock basis as
\begin{equation}
\rho_T=\sum_{n=0}^{\infty} \frac{\bar{n}^n}{\left(1+\bar{n}\right)^{n+1}}\ket{n}\bra{n}.
\end{equation}
The variance and skew information for the measurement of $ X_{\theta}$ in the state $\rho_T$  are calculated as $ I=\bar{n}+1/2-\sqrt{\bar{n}\left(\bar{n}+1\right)}$, and $\Delta^2 X_{\theta}=\bar{n}+1/2$, respectively. Therefore, the modified standard deviation, avoiding the classical mixing contribution, is $\bar{\Delta}X_{\theta}=1/\sqrt{2}$. Substituting the value of $\bar{\Delta}X_{\theta}$ instead of $\Delta X_{\theta}$ in equation~(\ref{Nonclassical_Area}), we calculated the nonclassical area of the thermal state $\sigma\left(\rho_T\right)-\sqrt{2}\,\pi$ as zero which in turn tell us that the thermal state is a classical state, as expected. The idea here is to remove the classical mixing part from the standard deviation in the measurement of rotated quadrature operator and make sure it contains only quantum uncertainty before performing the integration in equation~(\ref{Nonclassical_AreaPmode}). One can also perform the Monte-Carlo method to simulate the data from the homodyne detection measurements and obtain the results presented in this paper  \cite{Ariano1995,Ariano1998,Ariano1999,Ariano1999a}. 

\section{Conclusion}\label{sec6}
We have introduced a simple and easily computable nonclassicality indicator for the quantum states of light. The proposed nonclassicality indicator is based on
the standard deviation in the homodyne rotated quadrature operator’s measurement and is termed as {\it nonclassical area}. The nonclassical area can be viewed as an area projected by the optical tomogram of the state on the optical tomographic plane. If the nonclassical area spanned by the optical tomogram of a pure quantum state of light is nonzero, then the state is nonclassical. The nonclassical area is zero for the pure classical state (for a coherent state). A nonzero value of the nonclassical area for a given quantum state (except the state which are a probabilistic mixture of states) can be considered as a  sufficient condition for the state to be nonclassical. We have analyzed the nonclassical area spanned by the optical tomogram of several nonclassical states and observed that the essential features of nonclassicality shown by the states are consistently reflected in the nonclassical area indicator of nonclassicality. We have found that the nonclassical area associated with an arbitrary quantum state of light increases with the strength of any nonclassicality-inducing operations acting on the state, such as squeezing, photon addition, and superimposing two states. By investigating the effect of environment-induced decoherence on the nonclassical area, we have shown that the nonclassical area indicator of nonclassicality is robust against decoherence. The advantage of the nonclassical area measure is that the experimentalist can use equation~(\ref{p-modeNA}) to calculate the degree of nonclassicality associated with an arbitrary quantum state of light directly from the homodyne tomography data without using the density matrix or the quasiprobability distributions reconstructed from the optical tomogram. Thus, this method avoids the errors which may arise during the numerical method for reconstructing the density matrix or the quasiprobability distribution of the state.
\section*{Data availability statement}
All data that support the findings of this study are included within the article (and any supplementary files).
\section*{References}
\bibliographystyle{iopart-num}
\bibliography{reference}

\providecommand{\newblock}{}
\begin{thebibliography}{10}
\expandafter\ifx\csname url\endcsname\relax
  \def\url#1{{\tt #1}}\fi
\expandafter\ifx\csname urlprefix\endcsname\relax\def\urlprefix{URL }\fi
\providecommand{\eprint}[2][]{\url{#2}}

\bibitem{Glauber1963}
Glauber R~J 1963 {\em Phys. Rev.\/} {\bf 131} 2766--2788

\bibitem{Dodonov2002}
Dodonov V~V 2002 {\em J. Opt. B: Quantum Semiclass. Opt.\/} {\bf 4} R1

\bibitem{Sudarshan1963}
Sudarshan E~C~G 1963 {\em Phys. Rev. Lett.\/} {\bf 10} 277--279

\bibitem{Lvovsky2009}
Lvovsky A~I and Raymer M~G 2009 {\em Rev. Mod. Phys.\/} {\bf 81} 299--332

\bibitem{Mandel1979}
Mandel L 1979 {\em Opt. Lett.\/} {\bf 4} 205--207

\bibitem{Hillery1987}
Hillery M 1987 {\em Phys. Rev. A\/} {\bf 35} 725--732

\bibitem{Dodonov2000}
Dodonov V~V, Man'ko O~V, Man'ko V~I and W\"{u}nsche A 2000 {\em J. Mod. Opt.\/}
  {\bf 47} 633--654

\bibitem{Wunshe2001}
W\"{u}nsche A, Dodonov V~V, Man'ko O~V and Man'ko V~I 2001 {\em Fortschr.
  Phys.\/} {\bf 49} 1117--1122

\bibitem{Dodonov2003}
Dodonov V and Ren\'{o} M 2003 {\em Phys. Lett. A\/} {\bf 308} 249 -- 255

\bibitem{Marian2002}
Marian P, Marian T~A and Scutaru H 2002 {\em Phys. Rev. Lett.\/} {\bf 88}
  153601

\bibitem{Marian2004}
Marian P, Marian T~A and Scutaru H 2004 {\em Phys. Rev. A\/} {\bf 69} 022104

\bibitem{Nair2017}
Nair R 2017 {\em Phys. Rev. A\/} {\bf 95} 063835

\bibitem{Asboth2005}
Asb\'oth J~K, Calsamiglia J and Ritsch H 2005 {\em Phys. Rev. Lett.\/} {\bf 94}
  173602

\bibitem{Vidal2002}
Vidal G and Werner R~F 2002 {\em Phys. Rev. A\/} {\bf 65} 032314

\bibitem{Vogel2014}
Vogel W and Sperling J 2014 {\em Phys. Rev. A\/} {\bf 89} 052302

\bibitem{Kenfack2004}
Kenfack A and \.{Z}yczkowski K 2004 {\em J. Opt. B: Quantum Semiclass. Opt.\/}
  {\bf 6} 396

\bibitem{Hudson1974}
Hudson R 1974 {\em Rep. Math. Phys.\/} {\bf 6} 249--252

\bibitem{Lee1991}
Lee C~T 1991 {\em Phys. Rev. A\/} {\bf 44} R2775--R2778

\bibitem{Ryl2017}
Ryl S, Sperling J and Vogel W 2017 {\em Phys. Rev. A\/} {\bf 95} 053825

\bibitem{Gehrke2012}
Gehrke C, Sperling J and Vogel W 2012 {\em Phys. Rev. A\/} {\bf 86} 052118

\bibitem{Terhal2000}
Terhal B~M and Horodecki P 2000 {\em Phys. Rev. A\/} {\bf 61} 040301(R)

\bibitem{Sanpera2001}
Sanpera A, Bru\ss{} D and Lewenstein M 2001 {\em Phys. Rev. A\/} {\bf 63}
  050301(R)

\bibitem{Mraz2014}
Mraz M, Sperling J, Vogel W and Hage B 2014 {\em Phys. Rev. A\/} {\bf 90}
  033812

\bibitem{Vogel1989}
Vogel K and Risken H 1989 {\em Phys. Rev. A\/} {\bf 40} 2847--2849

\bibitem{Smithey1993}
Smithey D~T, Beck M, Raymer M~G and Faridani A 1993 {\em Phys. Rev. Lett.\/}
  {\bf 70} 1244--1247

\bibitem{Leonhardt1997}
Leonhardt U 1997 {\em Measuring the {Quantum} {State} of {Light}\/} (Cambridge:
  Cambridge University Press)

\bibitem{Bellini2012}
Bellini M, Coelho A~S, Filippov S~N, Man'ko V~I and Zavatta A 2012 {\em Phys.
  Rev. A\/} {\bf 85} 052129

\bibitem{Barnett1997}
Barnett S~M and Radmore P~M 1997 {\em Methods in theoretical quantum optics\/}
  (Oxford: Oxford University Press)

\bibitem{Rohith2015}
Rohith M and Sudheesh C 2015 {\em Phys. Rev. A\/} {\bf 92} 053828

\bibitem{Rohith2016}
Rohith M and Sudheesh C 2016 {\em J. Opt. Soc. Am. B\/} {\bf 33} 126--133

\bibitem{Sharmila2017}
Sharmila B, Saumitran K, Lakshmibala S and Balakrishnan V 2017 {\em J. Phys. B:
  At. Mol. Opt. Phys.\/} {\bf 50} 045501

\bibitem{Bazrafkan2003}
Bazrafkan M~R and Man’ko V~I 2003 {\em J. Opt. B: Quantum Semiclass. Opt.\/}
  {\bf 5} 357

\bibitem{Filippov2011}
Filippov S~N and Man'ko V~I 2011 {\em Phys. Scr.\/} {\bf 83} 058101

\bibitem{Korennoy2011}
Korennoy Y~A and Man'ko V~I 2011 {\em Phys. Rev. A\/} {\bf 83} 053817

\bibitem{Miranowicz2014}
Miranowicz A, Paprzycka M, Pathak A and Nori F 2014 {\em Phys. Rev. A\/} {\bf
  89} 033812

\bibitem{Gerry2005}
Gerry C and Knight P 2005 {\em Introductory quantum optics\/} (Cambridge
  university press)

\bibitem{Kral1990}
Král P 1990 {\em J. Mod.Opt.\/} {\bf 37} 889--917

\bibitem{Agarwal1991}
Agarwal G~S and Tara K 1991 {\em Phys. Rev. A\/} {\bf 43} 492--497

\bibitem{Zavatta2004}
Zavatta A, Viciani S and Bellini M 2004 {\em Science\/} {\bf 306} 660--662

\bibitem{Mancini1996}
Mancini S, Man'ko V and Tombesi P 1996 {\em Phys. Lett. A\/} {\bf 213} 1 -- 6

\bibitem{VanEnk2003}
van Enk S~J 2003 {\em Phys. Rev. Lett.\/} {\bf 91} 017902

\bibitem{Gardiner1991}
Gardiner C~W 1991 {\em {Quantum Noise}\/} (Berlin: Springer)

\bibitem{Biswas2007}
Biswas A and Agarwal G~S 2007 {\em Phys. Rev. A\/} {\bf 75} 032104

\bibitem{Sharmila2019}
Sharmila B, Lakshmibala S and Balakrishnan V 2019 {\em Quantum Inf. Process.\/}
  {\bf 18} 236

\bibitem{Hiroshima2001}
Hiroshima T 2001 {\em Phys. Rev. A\/} {\bf 63} 022305

\bibitem{Srinivasan1991}
Chaturvedi S and Srinivasan V 1991 {\em Phys. Rev. A\/} {\bf 43} 4054--4057

\bibitem{Luo2005}
Luo S 2005 {\em Phys. Rev. A\/} {\bf 72} 042110

\bibitem{Ariano1995}
D'Ariano G~M, Leonhardt U and Paul H 1995 {\em Phys. Rev. A\/} {\bf 52}
  R1801--R1804

\bibitem{Ariano1998}
D'Ariano G~M, Vasilyev M and Kumar P 1998 {\em Phys. Rev. A\/} {\bf 58}
  636--648

\bibitem{Ariano1999}
D'Ariano G~M, Sacchi M~F and Kumar P 1999 {\em Phys. Rev. A\/} {\bf 59}
  826--830

\bibitem{Ariano1999a}
D'Ariano G~M, Sacchi M~F and Kumar P 1999 {\em Phys. Rev. A\/} {\bf 61} 013806

\end{thebibliography}

\end{document}